\documentclass[ conference, article]{IEEEtran}

\IEEEoverridecommandlockouts
\usepackage{cite}
\usepackage{soul}
\usepackage{amsmath,amssymb,amsfonts}
\usepackage{graphicx}
\usepackage[switch]{lineno}
\usepackage{textcomp}
\usepackage{makecell}
\usepackage{xcolor}
\usepackage{tikz}
\usepackage{colortbl}
\usepackage{multirow}
\usepackage{multicol}
\usetikzlibrary{mindmap, shadows.blur}
\usepackage{comment}
\usepackage{amsmath}
\usepackage{mathtools, nccmath}
\usepackage{lipsum}
\usepackage{pifont,xspace}
\usepackage{algorithm}
\usepackage{algpseudocode}
\usepackage{color, colortbl}
\usepackage{subfig}
\usepackage{textcomp}
\usepackage{booktabs}
\usepackage{tabularx}
\usepackage{hyperref}
\usepackage{cleveref}
\usepackage{pifont}
\usepackage{url}
\usepackage{booktabs}
\usepackage{amssymb}
\usepackage{utfsym}
\usepackage{tcolorbox}

\DeclareUrlCommand\email{\urlstyle{tt}}

\newcommand*\circled[1]{\tikz[baseline=(char.base)]{\node[shape=circle,fill,inner sep=0pt,minimum size=1pt] (char) {\textcolor{white}{#1}};}}

\newcommand{\gps}[1]{\mathsf{#1}\xspace}

\newcommand{\pr}{\gps{PR}}
\newcommand{\pru}{\gps{PRU}}
\newcommand{\adr}{\gps{ADRng}}
\newcommand{\adru}{\gps{ADRngU}}
\newcommand{\cno}{\gps{CN0}}
\newcommand{\bbcno}{\gps{BbCN0}}
\newcommand{\agc}{\gps{Agc}}
\newcommand{\rsvtu}{\gps{RecSvTU}}
\newcommand{\state}{\gps{State}}

\newcommand{\cmark}{\textcolor{green}{\checkmark}} 
\newcommand{\xmark}{\textcolor{red}{\text{\sffamily X}}} 

\definecolor{GrayL}{gray}{0.95}
\definecolor{GrayM}{gray}{0.85}
\definecolor{GrayD}{gray}{0.65}

\def\BibTeX{{\rm B\kern-.05em{\sc i\kern-.025em b}\kern-.08em
    T\kern-.1667em\lower.7ex\hbox{E}\kern-.125emX}}
\begin{document}

\title{ \LARGE \bf \textit{AndroCon}: Conning Location Services in Android}
\author{
    \IEEEauthorblockN{\large Soham Nag}
    \IEEEauthorblockA{\large
        Center of Excellence in Cyber Systems \\
        and Information Assurance\\
        Indian Institute of Technology Delhi\\
        New Delhi, India\\
        jcs222663@csia.iitd.ac.in
    }
    \and
    \IEEEauthorblockN{\large Smruti R. Sarangi}
    \IEEEauthorblockA{\large
        Computer Science and Engineering\\
        Indian Institute of Technology Delhi\\
        New Delhi, India\\
        srsarangi@cse.iitd.ac.in
    }
}

\maketitle
\begin{abstract}
Ambient sensing, human activity recognition and indoor floor mapping are common targets
for attackers who hack mobile devices. Other than overt signals such as microphones and
cameras, other covert channels such as WiFi, Bluetooth and assisted GPS signal strengths have been
used to infer this information.
In the space of passive, receive-only sattelite GPS-based sensing, the state-of-the-art
was limited to using only the signal strength and location information up till now. 
This paper shows
that semi-processed GPS data (with 39  features)
that is now accessible to apps since the release of
Android 7 (with precise location permissions) can be used
as a strong leaky channel to sense the ambient, recognize human activity and map indoor
spaces with a very high accuracy (99\%+ in many cases). This paper presents the results
of a longitudinal study where semi-processed GPS measurements were taken over the course
of a year using different mobile devices spread out over a 40,000 sq. km geographical region.
Data was also collected on flights, cruise ships and high-altitude
locations. We thoroughly characterize all the satellite GPS signals and based on cross-correlation
analysis extract the best set of features that preserve vital information.

We propose a novel method, AndroCon, that comprises linear discriminant
analysis, unscented Kalman filtering, gradient boosting and random forest based learning
to yield a highly accurate ambient and human activity sensor. AndroCon relies on simple
ML algorithms that can run surreptitiously and provide partially explainable results. 
We easily
identify difficult scenarios such as being inside a metro, a person waving a hand in front
of the mobile device, being in front of a staircase
and the presence of people in the room (not necessarily
holding mobile phones). 
This is the most comprehensive study on satellite GPS-based sensing till date.
\end{abstract}
\section{Introduction}

The integration of advanced wireless technologies such as WiFi, 4G, and 5G along with precise-positioning technologies
such as the Global Positioning System (GPS) in mobile devices has fuelled the growth of a
plethora of applications that use location-based services (LBS)
~\cite{8377571,lbs}. The market for such applications is currently 50 billion USD and
is anticipated to soar to 400 billion USD by 2030 (CAGR of 24.6\%)~\cite{Jackiewicz2023, alliedmarket}.

LBS services are used by applications that provide the following services:
navigation, local search, traffic alerts, weather updates, home delivery,
ride-sharing and device
tracking -- they enhance the user experience significantly.
As of today, such applications have become indispensable. 94\% of smartphone users
rely on them daily and 84\% of small to medium-sized businesses are reported to
have seen a surge in the footfall due to location-based
marketing~\cite{lbsstatics}.

We shall show in this paper that while LBSs offer a lot of convenience, they pose significant privacy risks -- many of
them were hitherto unknown and undiscovered. Let us first look at the known privacy risks of the {\em precise location signal}.
It conveys location information, which for obvious reasons 
leaks privacy information~\cite{wang2018privacy}. Our claim in this paper is that GPS signals that are used to find the location can be used for other surreptitious 
activities as well: sensing the ambient, tracking human activity, and finding more about the
layout (floorplan). This can be done because modern GPS chips share a lot of information with applications such as the
signal strength, Doppler shift, signal-to-noise ratio (SNR), etc. Modules in the GPS chip use this information to deduce the precise location ($x$, $y$, and $z$ coordinates) of the device. Our 
claim in this paper is that we can 
find alternative uses for this information that is readily available at the application level.
This will allow us to sense the user's environment. Using such semi-processed GPS signals -- already provided by all GPS chips to Android applications --
as an ambient sensor is entirely novel (to the best of our knowledge). 

There is some work in the field of ambient sensing using other modalities. 
For example, 
smart speakers~\cite{lau2018alexa} and WiFi signals\cite{atallah2007behaviour,carrera2016real} have been used to find out details of the users' ambience. 
In the former case, background sounds in the environment have been used to guess the
nature of the user's surroundings and in the latter case, the WiFi signal strengths have
been used as an information-leaking channel. Similar work exists with other more overt channels such
as mobile phone cameras~\cite{kang2020review, beddiar2020vision,yadav2021review}. 
Bluetooth beacons emit signals that can be picked up by nearby devices.
This allows attackers to infer proximity and movement patterns of users~\cite{bibbo2022overview}.
Such ambient 
information is quite beneficial. One of the innocuous uses is that it allows third parties
to influence online shopping behavior through
targeted advertising~\cite{ADambient}. However, there are more pernicious applications 
as well that pose genuine privacy risks such as figuring out that a person of interest
is having a meeting in a small room, when he was actually supposed to be there at a party
10 meters away. The basic cell phone location information will not be able to discriminate
between these two situations, whereas our solution \textit{AndroCon} will be able to. 

Let us further elaborate on our use case. Assume that somehow a user was conned into
installing an Android app. She further gave it precise location permissions. Almost all e-commerce, ride-sharing apps and some games like Pokemon Go~\cite{pokemonGo} necessitate such
permissions. Now, our claim is that even if the device is in flight mode and
all communication
channels are off (WiFi, mobile data, NFC, Bluetooth), if we can still continue to read semi-processed GPS data, then an important information-leaking side-channel exists. The GPS data can be used to figure
out important information about the environment that includes (but is not limited to)
whether the user is in a closed space or open space, is the place crowded, is the user sitting or standing, moving quickly or slowly, underground or overground, inside a flight, within a lift
or close to a staircase (refer to Section~\ref{eval} for the full list). There is no need to take the help of any other kind of sensor such as 
the camera, accelerometer or microphone. Our claim is that semi-processed GPS data (in pure receive-only mode) is sufficient. In this paper, we shall advance various arguments to 
convince the reader that this is indeed possible. We shall make theoretical arguments,
show the results of simulations, show real measurement data using GPS sensors and finally
show extensive evaluation results using 5 phones collected over a period of 1 year. 

In this paper, we shall demonstrate the potential of semi-processed GPS parameters,
both independently and collectively, to
discern user activity (both static and dynamic) and the
ambience in diverse settings (crowded, open area, indoor, metro tunnel, flight). 
As GPS signals bounce off the user and their surroundings, 
they capture the environmental context, generating a unique pattern for each
setting~\cite{pu2013whole, gu2015paws, sekiguchi2021phased}. These details are intrinsic to semi-processed GPS parameters, which we fingerprint for the purpose of
ambient sensing. However, GPS signals, are
inherently noisy due to multipath effects and interference with other signals~\cite{yuan2020gps} --
this necessitates data cleaning prior to the fingerprinting exercise. We employ a nonlinear
Kalman filter for noise filtering. It preserves the essential signal variations that our algorithm
needs~\cite{hu2020unscented}. Subsequently, we apply Linear Discriminant Analysis (LDA) for 
feature reduction and identify linear combinations of semi-processed parameters that effectively distinguish between different ambient contexts. 
The resulting data is then fed into an ML model for ambient and activity classification. 
Additionally, we use the semi-processed parameters and GPS signal strength, in conjunction with user trajectories and graph optimization (GO) techniques, to detect the indoor layout of the user's location.

This paper makes the following contributions:

\circled{1} To the best of our knowledge, this is the first study that characterizes in-depth the potential of semi-processed GPS signal for ambient sensing.

\circled{2} We show how apps using GPS sensing can 
covertly capture semi-processed GPS data without consent and utilize it to discern the user's ambience (99.6\%) and activities (87\%) across
{\em diverse} settings, effectively
jeopardizing privacy. At the moment, this vulnerability affects {\em 90\%} of Android users.

\circled{3} Evaluation of
ML models to effectively classify different user activities and ambient settings using 8 semi-processed GPS
parameters as inputs. 

\circled{4} The ability of attackers to infer floor maps/indoor layouts (error margin of 4 meters), using semi-processed GPS data and user trajectories, without needing access to other ambient
information leaking sources such as cameras.

The structure of this paper is as follows: \cref{sec:background} provides the necessary background information. In
\cref{sec:rawgps}, we discuss the semi-processed GPS parameters that we use
and their effectiveness in classifying activities and
environments followed by characterization and validation of semi-processed data readings in \cref{sec:emperical}. The
architecture of the model is shown in \cref{model} and its performance is evaluated in \cref{eval}. The overview and
evaluation of the layout estimation
algorithm are presented in \cref{layout}.  We review the relevant literature in \cref{related}, and
finally, conclude in \cref{sec:conclusion}.

\section{Background}
\label{sec:background}

\subsection{Overview of GPS}
\label{sec:gps}

The Global Positioning System (GPS) was established by the US Department of Defense in 
1973. It is 
a member of the Global Navigation Satellite System (GNSS) constellation.
GPS, in 2024, 
has 31 satellites that orbit the Earth in six planes of rotation. All of them
are in the medium earth
orbit (20,000 kms above sea level 
~\cite{gpsworking}). These satellites transmit navigation signals, which
contain 
precise information about their position, velocity and current time (PVT).
This information is generally resistant to 
inclement weather conditions. To accurately calculate the geographical location, 
a GPS receiver requires signals from at least four satellites.

Let us explain the physics. Every GPS satellite has an atomic clock that very accurately
maintains the time. The uncertainty is less than 1 part in $10^{16}$. Assume that
a message is sent from a satellite at time $t_s$ and it is received at time $t_r$. 
Then the distance $d$ between the sending satellite and the receiver is shown in Equation~\ref{eqn:tstr}.

\begin{equation}
\label{eqn:tstr}
    d = (t_s - t_r + \Delta) \times c
\end{equation}

Here, $c$ is the speed of light and $\Delta$ is the clock skew between the non-ideal receiver clock
and the ideal sender clock.  
Now, the Euclidean distance $d$ is also give by another equation (see Equation~\ref{eqn:edist}).

\begin{equation}
\label{eqn:edist}
    d = \sqrt{(x_s - x_r)^2 + (y_s - y_r)^2 + (z_s - z_r)^2}
\end{equation}

Here, $\langle x_s, y_s, z_s \rangle$ are the coordinates of the satellite and $\langle x_r, y_r, z_r \rangle$ are
the coordinates of the receiver. 
There are four unknowns here: three of them are the coordinates of the receiver and the fourth unknown is the
clock skew $\Delta$. For four unknowns, we need four equations. We thus need data from four satellites.

We need to note that Equation~\ref{eqn:tstr} is an ideal relationship. 
There are delays induced due to the ionosphere, signal interference, multipath effects, etc. Hence, modern
receivers need to apply many different correction factors. The actual relationship is thus
more complex.
Hence, there is a need to also transmit many other parameters (described in Section~\ref{sec:rawgps}) as well
such that receivers can correct their data. These include the pseudorange along with SNR, phase shifts, Doppler shifts
and details about
the satellite and the constellation (see Table~\ref{tab:gnssmeasurement}). A GPS message also contains 
other parameters about the receiver's clock (other than the time itself).

\subsubsection{GPS pipeline}

\begin{figure*}[htb]
\centering
\vspace*{-6.5ex}
    \includegraphics[scale=0.65]{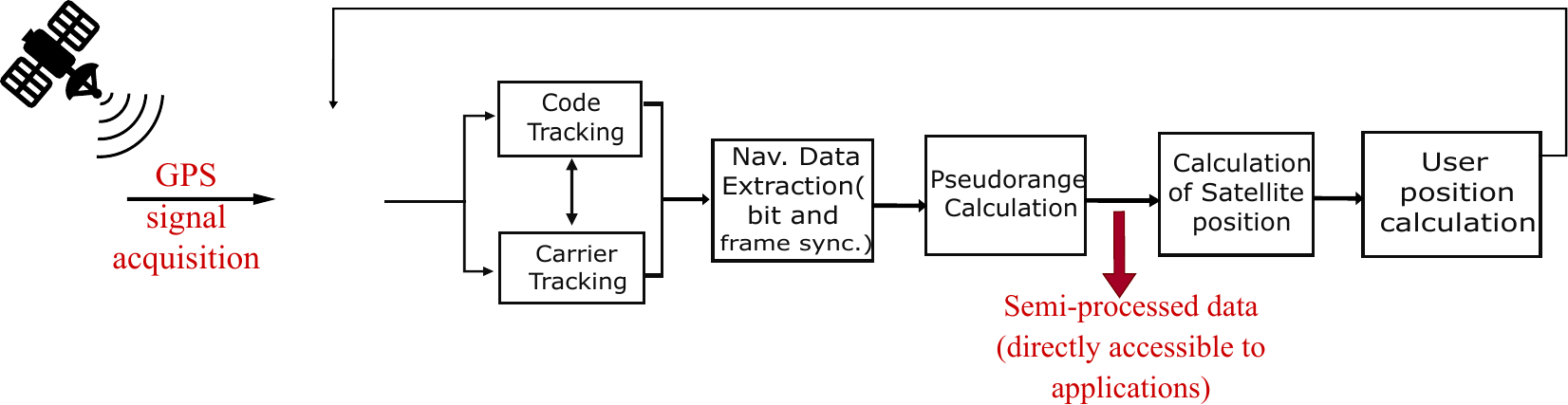}
            \caption{GPS pipeline: From signal acquisition to position calculation}
    \label{fig:gpspipeline}
\end{figure*}

Figure~\ref{fig:gpspipeline} shows the GPS processing pipeline at the receiver.
Each GPS satellite uses a unique
pseudorandom code ($PRN$) to modulate its signal for unique identification, ensuring security and
enhancing the robustness~\cite{vu2014code}. Code and carrier tracking loops monitor all such incoming
GPS signals, demodulate and identify them. The code tracking
loop compares the incoming signal's code with the codes of one of the replicas. 
Through cross-correlation analysis, the receiver
finds the code (and satellite) corresponding to the signal.
This aids in calculating
the pseudorange -- apparent distance between the satellite and receiver~\cite{vu2014code} --
which is one of the most important parameters of interest for distance calculation. 

Modern GPS chipsets allow an application to tap the information at any point in the GPS
processing pipeline. In general, applications do not tap the information available at the
intermediate stages because they don't find it
useful. Only the output of the final stage is used by applications -- it is the device's precise
location information. We decided to tap the pipeline in the middle and extract 9 parameters
of
interest out of 32 parameters 
(refer to Table~\ref{tab:gnssmeasurement}). Hence, we refer to these parameters as
{\em semi-processed GPS data}. 

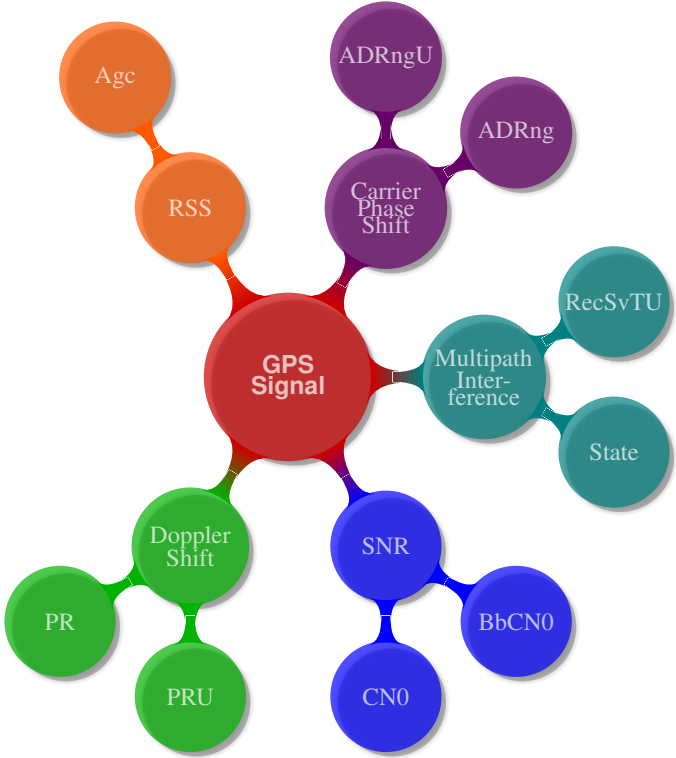
\begin{figure}[!htb]
\centering
\begin{tikzpicture}[mindmap, concept color=red!80!black, font=\sf\bfseries\large, text=white, grow cyclic,
    every node/.style={concept, scale=0.55, text width=2.5cm, drop shadow={shadow xshift=0.2ex, shadow yshift=-0.2ex, blur shadow={shadow blur steps=5}}, opacity=0.7}, level 1/.append style={level distance=2.6cm}, level 2/.append style={level distance=2.0cm}]
  \node[concept] {\LARGE{GPS Signal}}
    child[concept color=green!70!black] { node {\LARGE Doppler Shift}
      child { node {\LARGE PR} }
      child { node {\LARGE PRU} }
    }
    child[concept color=blue] { node {\LARGE SNR}
      child { node {\LARGE CN0} }
      child { node {\LARGE BbCN0} }
    }
    child[concept color=teal] { node {\LARGE Multipath Interference} 
      child { node {\LARGE State} }
      child { node {\LARGE RecSvTU} }
    }
    child[concept color=violet!80!black] { node {\LARGE Carrier Phase Shift} 
      child { node {\LARGE ADRng} }
      child { node {\LARGE ADRngU} }
    }
    child[concept color=orange!70!red] { node {\LARGE RSS}
      child { node {\LARGE Agc} }
    };
\end{tikzpicture}
\caption{Representation of the semi-processed GPS signal parameters that we use. RSS is ``Received Signal Strength''}
\label{fig:mindmap}
\vspace{-2ex}
\end{figure}

A brief taxonomy of these parameters is presented in Figure~\ref{fig:mindmap}. We broadly
use five types of parameters that can be classified into the following buckets: 
received signal power, carrier phase shift, multipath inteference, signal-to-noise ratio (SNR)
and the Doppler shift (due to satellite motion). These parameters can be used to extract
information about the surroundings because they are influenced by it. For instance, if
there are a lot of objects around the receiver, the multi-path interference and SNR will be high.

\begin{table*}[htbp]
    \caption{List of semi-processed GPS signal parameters \label{tab:gnssmeasurement}}
    
    \begin{tabular}{|l|l|p{0.48\textwidth}|l|}
        \hline
        \rowcolor{GrayD}
        {\bf Field} & {\bf Notations} & {\bf Description} & {\bf Unit} \\
        \hline
        \hline

         \small $PseudorangeRate$ &  $\pr$ & \small  Pseudorange rate at the associated timestamp. & \small m/s \\
        \hline
         \rowcolor{GrayM}\small $PseudorangeRateUncertainty$ &  $\pru$ & \small Pseudorange's rate uncertainty (1-$\sigma$). & \small m/s \\
        \hline 
         \small $ReceivedSvTimeUncertainty$ &  $\rsvtu$ & \small Error estimate (1-$\sigma$) for the received GNSS time. & \small ns \\
        \hline
        \rowcolor{GrayM}\small $AccumulatedDeltaRange$ &  $\adr$ & \small Accumulated delta range since the last channel reset. & \small m \\
        \hline
       \small $AccumulatedDeltaRangeUncertainty$ &  $\adru$ & \small Uncertainty of the accumulated delta range (1-$\sigma$). & \small m \\
        \hline
        \rowcolor{GrayM} \small $\gps{C}\gps{N}0$ &  $\cno$ & \small Carrier-to-noise density in the range [0,63]. & \small dB-Hz \\
        \hline
         \small $BasebandCn0DbHz$ &  $\bbcno$ & \small Baseband carrier-to-noise density (Added in API level 30). & \small dB-Hz \\
        \hline
         \rowcolor{GrayM}\small $AgcDb$ &  $\agc$ & \small Incoming signal power. & \small dB \\
        \hline
        \small $State$ &  $\gps{State}$ & \small Integer representing satellite sync state, with each bit indicating a specific measurement status. & - \\
        \hline
    \end{tabular}
\end{table*}

\subsection{ Location Accuracy Permissions in Android}
In Android, the location accuracy permission, facilitated by 
{\small $\mathtt{ACCESS\_COARSE\_LOCATION}$} and {\small $\mathtt{ACCESS\_FINE\_LOCATION}$}, 
enables apps to specify the desired precision to access the location of the device. 
{\small $\mathtt{ACCESS\_COARSE\_LOCATION}$} utilizes cell tower or Wi-Fi network
data for positioning, offering 100 meters accuracy~\cite{finelocaccuracy}, suitable for weather or news apps. {\small $\mathtt{ACCESS\_FINE\_LOCATION}$} 
provides higher precision through GPS-based positioning methods,
typically within 50 meters, occasionally as fine as 3 meters or 
better~\cite{Buildlocation-awareapps}, crucial for mapping or navigation apps. 
In Android 12 and beyond, apps must request both permissions, irrespective of 
their precision requirements, to suit user preferences. Notably, accessing semi-processed 
GPS measurements requires {\small $\mathtt{ACCESS\_FINE\_LOCATION}$} permission.

\subsection{Linear Discriminant Analysis (LDA)}
Any kind of learning that involves a lot of parameters tends to be inefficient unless we have a large number
of training examples. Hence, dimensionality reduction needs to be done. We use
LDA, which is a widely used supervised-learning technique for dimensionality reduction in pattern-classification 
applications~\cite{sugiyama2006local, sarveniazi2014actual, sugiyama2007dimensionality}.
It projects a dataset into a lower-dimensional space that enhances {\em class separability}
through a linear combination of features.
This is done to
address the curse of dimensionality and reduce
computational costs~\cite{verleysen2005curse}. Unlike a cognate technique namely Principal Component 
Analysis (PCA), which identifies directions 
of maximum variance, LDA creates
a transformation that maximizes the \underline{separation} between different classes. 

LDA operates by finding a projection matrix that transforms the 
original data into a new space where samples from different classes are {\em more 
distinct}. This is achieved by maximizing the between-class variance while 
minimizing the within-class variance. The projection matrix is determined 
through an optimization process involving scatter matrices that represent variances. 
This involves solving an eigenvalue problem, where the eigenvectors form the projection matrix.
By transforming the data in this manner, LDA facilitates a more effective 
and efficient classification, making it a valuable tool 
for the exploitation of the semi-processed GPS data in our research.

\subsection{Unscented Kalman Filter}
The Unscented Kalman Filter (UKF) is an advanced unsupervised
technique used for noise filtering in \textit{nonlinear} systems. It
predicts the future state of a nonlinear system
and updates this prediction using incoming noisy measurements.
Unlike the traditional Kalman Filter, which assumes linearity, the UKF employs a deterministic sampling approach to
generate sample points around the mean, which are then passed as inputs to nonlinear functions. This method accurately
captures the mean and covariance accurately, enhancing its effectiveness in complex applications. The UKF is
widely used in fields such as robotics, and navigation, where precise estimation in the presence of noise is
crucial.

Consider a nonlinear state-space model at time $t$:

\begin{equation}
x_t = f(x_{t-1}) + w_t
\end{equation}

\begin{equation}
z_t = h(x_t) + v_t
\end{equation}

$x_t$ is the state vector and $z_t$ is the 
corresponding noisy measurement at time $t$;  $f(\cdot)$ is the state transition function;  $h(\cdot)$ is the observation function; $w_t \sim \gps{N}(0, Q)$ is the process noise representing the uncertainty in the model's state transition with covariance $Q$; $v_t \sim \gps{N}(0, R)$ is the measurement noise representing the uncertainty in the measurements with covariance $R$.

The Unscented Transform approximates the state distribution using a set of sigma points. These points are passed as inputs to the the nonlinear functions $f$ and $h$. 

\noindent{{\emph {\circled{1} Sigma Point Generation}}}

Sigma points $\chi_{t-1}^{(i)}$ are generated around the mean $\hat{x}_{t-1}$:

\begin{equation}
\chi_{t-1}^{(0)} = \hat{x}_{t-1}
\end{equation}

\begin{equation}
\chi_{t-1}^{(i)} = \hat{x}_{t-1} + \sqrt{(L + \lambda)P_{t-1}}, \quad i = 1, \ldots, L
\end{equation}

\begin{equation}
\chi_{t-1}^{(i+L)} = \hat{x}_{t-1} - \sqrt{(L + \lambda)P_{t-1}}, \quad i = 1, \ldots, L
\end{equation}

${P}_{t-1}$ is the error covariance matrix; $L$ is the dimension of the state vector, and $\lambda$ is a scaling parameter describing the spread of the sigma
points around $\hat{x}_{t-1}$. Note that in the first case we are adding a scaled version of the error, and in the second case we are subtract it.

\noindent{ {\emph {\circled{2} Prediction Step}}}

The sigma points are next passed to the state transition function to generate a new set of samples.

\begin{equation}
\chi_{t|t-1}^{(i)} = f(\chi_{t-1}^{(i)}) \quad   ( i= 0,1,2,\ldots,2L)
\end{equation}

The predicted mean and covariance are computed as:

\begin{equation}
\hat{x}_{t|t-1} = \sum_{i=0}^{2L} W_{m}^{(i)} \chi_{t|t-1}^{(i)}
\end{equation}

\begin{multline}
P_{t|t-1} = \sum_{i=0}^{2L} W_{c}^{(i)} \left[\chi_{t|t-1}^{(i)} - \hat{x}_{t|t-1}\right] \\
\times \left[\chi_{t|t-1}^{(i)} - \hat{x}_{t|t-1}\right]^T + Q
\end{multline}

\[
\begin{cases}
W_{m}^{(i)} = 1- \frac{1}{\lambda^2}i = 0 \\ 
W_{c}^{(i)} = \frac{1}{2L\lambda^2}i  = 1, 2, \ldots, 2L
\end{cases}
\]
$W_{m}^{(i)}$ and $W_{c}^{(i)}$ are the weights for the mean and covariance, respectively. 

\noindent{ {\emph {\circled{2} Update Step}}}

The predicted measurements are obtained by passing the sigma points to the observation function ($h$).

\begin{equation}
\gamma_{t|t-1}^{(i)} = h(\chi_{t|t-1}^{(i)})
\end{equation}

The predicted measurement's mean and covariance are as follows:

\begin{equation}
\hat{z}_{t|t-1} = \sum_{i=0}^{2L} W_{m}^{(i)} \gamma_{t|t-1}^{(i)}
\end{equation}

\begin{equation}
S_t = \sum_{i=0}^{2L} W_{c}^{(i)} \left[\gamma_{t|t-1}^{(i)} - \hat{z}_{t|t-1}\right] \left[\gamma_{t|t-1}^{(i)} - \hat{z}_{k|t-1}\right]^T + R
\end{equation}

\begin{equation}
C_{t} = \sum_{i=0}^{2L} W_{c}^{(i)} \left[\chi_{t|t-1}^{(i)} - \hat{x}_{t|t-1}\right] \left[\gamma_{t|t-1}^{(i)} - \hat{z}_{t|t-1}\right]^T
\end{equation}

${C}_{t}$ is the cross-covariance matrix that represents the covariance between the state and the sigma points. 
${S}_{t}$ is the residual covariance matrix that includes the measurement noise covariance.  The Kalman gain $K_{t}$ and
the updated state and covariance are as follows:

\begin{equation}
K_{t} = C_{t} S_{t}^{-1}
\end{equation}

\begin{equation}
\hat{x}_t = \hat{x}_{t|t-1} + K_t (z_t - \hat{z}_{t|t-1})
\end{equation}

\begin{equation}
P_{t} = P_{t|t-1} - K_{t} S_{t} K_{t}^T
\end{equation}

We keep repeating this process
from the sigma point generation step for the 
subsequent inputs until all the data is processed.
\section{The Value of Semi-processed GPS Data}
\label{sec:rawgps}

In this section, we begin by examining the way semi-processed GPS
data is retrieved on an Android platform. 
Subsequently, we explain all the 9 GPS parameters, and finally
explain how ambient sensing can be achieved.

\subsection{Accessing Semi-Processed GPS Measurements}
Prior to Android 7, developers accessed location details via the {\url{android.gsm.location}} API, which provided basic
satellite information like C/N0 (carrier-to-noise ratio), azimuth and elevation, along with the fundamental National Marine Electronics
Association (NMEA) sentences containing the PVT (position, velocity and time) solution~\cite{zangenehnejad2021gnss}. 
Multipath effects and potential interference greatly degrade the positioning accuracy by altering the true distance
between satellites and the user; this causes pseudorange errors. GPS receivers attempt to correct these errors, but
since these measurements aren't directly accessible to users, they rely on often inadequate receiver-embedded correction
models~\cite{filic2018smartphone}. To foster the development of improved correction models for higher accuracy, Google
has made these details public on Android phones.
This information comprises the semi-processed GPS data, which is available to developers via
the {\url{android.location.GnssMeasurements}} API within the {\url{android.location}} package (starting from Android
7~\cite{Malkos2016}).

\subsection{Information Present in Semi-Processed GPS Measurements}
\label{sec:exploit}

Even though similar problems have been solved in the case of Wi-Fi and Bluetooth signals,
the same techniques cannot be reused. We need to devise novel methods to select features, 
identify the nature of the noise, clean the data, reduce the dimensionality, choose the settings
and ML models of interest and comprehensively evaluate the design. Let us 
understand in more detail the 9 GPS parameters of interest. The rest 23 (out of 32)
parameters were eliminated because they were not recorded on all the phones (variability across
chipsets), the values were quite unstable or were of no use (such as the details of the satellite). 

\noindent{\textbf {\emph {\circled{1} Doppler Shift }}}:

The Doppler shift indicates a change in frequency due to the relative motion between the
signal source and the receiver. Given that the satellites are moving and the receiver
itself may be non-stationary, a Doppler shift is expected. 
A positive shift indicates that the satellite
is moving towards the receiver -- this increases the perceived
signal frequency. Conversely, 
a negative shift occurs as the satellite moves away, effectively
decreasing the frequency. 

\begin{equation}
\begin{aligned}
\pr = \textit{-k} \times \textit{dopplershift} \text{ (where \textit{k} is a constant)}  
\end{aligned}
\end{equation}

This metric is significant in discerning the following scenarios.

\textbf{\textit{Identifying Motion State:}} A high value of $\pr$ suggests movement, while a low or near-zero value indicates a stationary state or slow movement. 

\textbf{\textit{Distinguishing Environmental Context:}} Multipath phenomena, 
where signals reflect off surfaces before reaching the receiver affect Doppler shifts. 
In open areas with minimal multipath effects, the $\pr$ measurements remain stable, while in crowded or indoor areas, they can fluctuate significantly due to multipath effects. 

Similarly, $\pru$ quantifies the uncertainty in $\pr$  measurements, which varies with environmental conditions. In crowded or mobile environments, multipath effects and irregular Doppler measurements increase this uncertainty, resulting in high $\pru$ values. In open areas with minimal multipath effects, the uncertainty is lower, leading to low $\pru$ values.
This analysis underscores the utility of $\pr$ and $\pru$ in inferring both the state of
motion and the environmental context of individuals based on semi-processed GPS measurements.

\noindent{\textbf {\emph {\circled{2} Carrier Phase}}: 

This quantifies the accumulated error in the distance between the satellite and the receiver. It is calculated by measuring the  
phase shift.  A positive value indicates that the satellite
vehicle ($SV$) is moving away from the receiver, while a negative value
signifies that the $SV$ is moving towards the receiver.

\begin{equation} \label{eq1}
\begin{aligned}
\adr  = -k * carrier phase \,\,
\text{(where \textit{k} is a constant)}  
\end{aligned}
\end{equation}
This facilitates the differentiation between the following scenarios. 

\textbf{\textit{Identifying Motion State:}} A high change in the $\adr$ suggests motion, while a low or near-zero change
indicates stationary or slow movement.

\textbf{\textit{Distinguishing Environmental Context:}} In open areas with minimal multipath interference, $\adr$ values are stable. Conversely, in crowded or indoor environments, reflections cause phase shifts, altering the carrier phase and resulting in variable $\adr$ measurements.

Similarly to $\pru$, $\adru$ reflects the uncertainty in $\adr$ measurements.

\noindent{\textbf {\emph {\circled{3} Signal To Noise Ratio}}:

$\cno$, or Carrier-to-Noise (C/N) density in dB-Hz, is a crucial metric for evaluating GNSS signal quality. It
typically ranges from 10 to 50 dB-Hz with a potential spectrum extending from 0 to 63 dB-Hz in exceptional circumstances~\cite{gnssmeasurement}. Variances in $\cno$ allow differentiation between the following environmental conditions.

In congested or enclosed areas with signal interference, $\cno$ levels decrease, indicating heightened noise in the GPS signal and suggesting the individual's presence in such environments.
Conversely, open-sky scenarios with minimal interference increase $\cno$ levels, signifying reduced noise in the GPS signal and implying the individual's presence in open-sky or less congested areas.

Similarly, $\bbcno$ represents the C/N of the signal at the baseband, obtained after demodulating the GPS signal received by the antenna. This value is typically slightly weaker than the C/N measured at the antenna port ($\cno$)~\cite{gnssmeasurement}.
Our experiments demonstrated a strong correlation coefficient(0.98) between $\cno$ and $\bbcno$, indicating similar behavior in distinguishing environments/activities.

\noindent{\textbf {\emph {\circled{4} Received Signal Strength (RSS)}}:

The $\agc$ acts as a dynamic amplifier, adjusting the power of incoming signals. Negative or low $\agc$ value indicates potential interference or jamming~\cite{gnssmeasurement}. Notably, this value remains consistent under the same level of incoming signal power, helping differentiate user activity (motion/rest).

Furthermore, $\agc$ can distinguish between different ambient conditions, such as crowded urban areas or open spaces near electromagnetic transmission towers. In highly interfered environments like crowded areas, $\agc$ levels are low, while in less congested or minimally interfered spaces, $\agc$ values are higher.

\noindent{\textbf {\emph {\circled{5} Multipath Interference}}:

The \textit{State} field indicates the current synchronization state for each satellite signal.  It can assume a value of either 0 or a combination of different synchronization states~\cite{gnssmeasurement}.

When the \textit{State} is {\footnotesize \url{STATE_MSEC_AMBIGUOUS}} (value: 16), it implies millisecond-level ambiguity in the GPS measurement's tracking state due to multipath effects~\cite{gnssmeasurement}, indicating a congested or enclosed environment.

$\rsvtu$ represents the error estimate (1-$\sigma$) for the received GPS time of a particular $SV$, typically influenced by multipath interference. As it quantifies uncertainty, its effectiveness in distinguishing user activity and environmental conditions mirrors that of $\pru$ And $\adru$.
\section{Empirical study of the semi-processed GPS signal characteristics}
\label{sec:emperical}
This section aims to establish a mapping between EM wave parameters, such as RSS and Doppler shift, which are utilized for ambient sensing, and the corresponding semi-processed GPS parameters. Subsequently, it examines the correlation among semi-processed parameters to reinforce the idea that they can be \textit{fused} to achieve improved classification outcomes.

\subsection{Real-World Data Collection}
Semi-processed GPS data was logged using the GnssLogger~\cite{gnsslogger} Android application on five different Android phones (refer to Table~\ref{tab:phone}).
Note that phones that use the Samsung Snapdragon chipset
do not allow the user to retrieve the $\adr$ measurements~\cite{rawgnssandoid}. 
We deliberately used two phones with the same version (Redmi Note 9 Pro Max) to characterize 
variations across two phones of the same model. There was some variation even in this case. 
Therefore, noise filtering
is required to ensure that our technique works across phones in a robust manner.

\textbf{\begin{table}[htb]
\caption{\small Details of the phones \label{tab:phone}}
    \resizebox{0.48\textwidth}{!}{
\centering
\begin{tabular}{|>{\centering\arraybackslash\small}p{2cm}|>{\centering\arraybackslash\small}p{2cm}|>{\centering\arraybackslash\small}m{2cm}|>{\centering\arraybackslash\small}m{2cm}|}
\hline
\rowcolor{GrayD}
\textbf{Model Name} & \textbf{Android version} & \textbf{Chipset} & \textbf{Year}\\
\hline
 Redmi Note 9 Pro Max & Android 11 & \textit{Snapdragon 720G SoC} & 2020 \\ \hline
\rowcolor{GrayM} Redmi Note 9 Pro Max & Android 11 & 
 \textit{Snapdragon 720G SoC} & 2021 \\ \hline
Galaxy A54 & Android 14 & \textit{Exynos 1380}
 &2023 \\ \hline
\rowcolor{GrayM} OnePlus Nord CE2 & Android 13 & \textit{MediaTek Dimensity 900} & 2022 \\ \hline
 Redmi Note K20 Pro & Android 12 & \textit{Snapdragon 855} & 2020 \\ \hline
\end{tabular}
}
\end{table}}

\subsection{Datasets: Synthetic (Kaggle) and Real-World}
Along with the data that we collect using these phones, we use open-source datasets as well. This is needed to 
ensure that similar effects are also being seen in those datasets and our noise filtering techniques work.
We utilized the Kaggle GNSS dataset~\cite{kagglegnss}; it contains 39 traces collected using
the Pixel 4, Pixel 4 XL and Xiaomi Mi8 phones (resp.). 
This dataset primarily contains measurements collected during motion. 
To simulate diverse user activities and environments, 
we generated supplementary data at different sites using our 5 phones.
The settings include an open ground, indoors, while standing and sitting. 
The data collection process lasted approximately 45-150 minutes at each site on an average.

The ionosphere's electron density variations cause fluctuations in GPS signals~\cite{dubey2006ionospheric}, resulting in measurement variability. Furthermore, diverse weather conditions can introduce additional fluctuations.
This leads to erroneous measurements and potential misclassification.
To simulate real-world settings and mitigate these effects, semi-processed data was collected three times 
over the course of a year under various atmospheric conditions and at different times of the day.
We tried to minimize the experimental noise by ensuring the ambient setting was the same. Of course, there
were variations due to the weather, position of satellites, etc. We did not try to achieve consistency
in these parameters.

\subsection{Calibration with Real Sensors: Signal Power}
Researchers have primarily used the RSS, the Doppler shift and SNR for designing
human activity recognition (HAR) and ambient sensing systems for WiFi, Bluetooth and cellular
tower signals ~\cite{pu2013whole, gu2015paws, sekiguchi2021phased}. We, of course, use many
more parameters.

However, we stick to these three parameters for the purpose of characterization and calibration of our setup. 
We map these parameters to the corresponding semi-processed GPS parameters. 
The GPS signal inherently captures the $\agc$ (RSS or signal power). We used
an RF Explorer Spectrum Analyzer (Model B34J7ML7J58J9MD6)~\cite{RFExplorerSpectrumAnalyzer}, which records
the signal RSS. We set the RF Explorer to the L1 GPS band frequency range (1575.42
MHz)~\cite{TimeandFrequencyNIST2023}.
We logged the GPS's RSS using the measurement device (RF Explorer) 
and simultaneously, we logged the $\agc$. We found a linear relationship between them (refer to
Figure~\ref{fig:rssicorr}).

\begin{figure}[h]
\begin{center}
    \includegraphics[width=1.1\columnwidth]{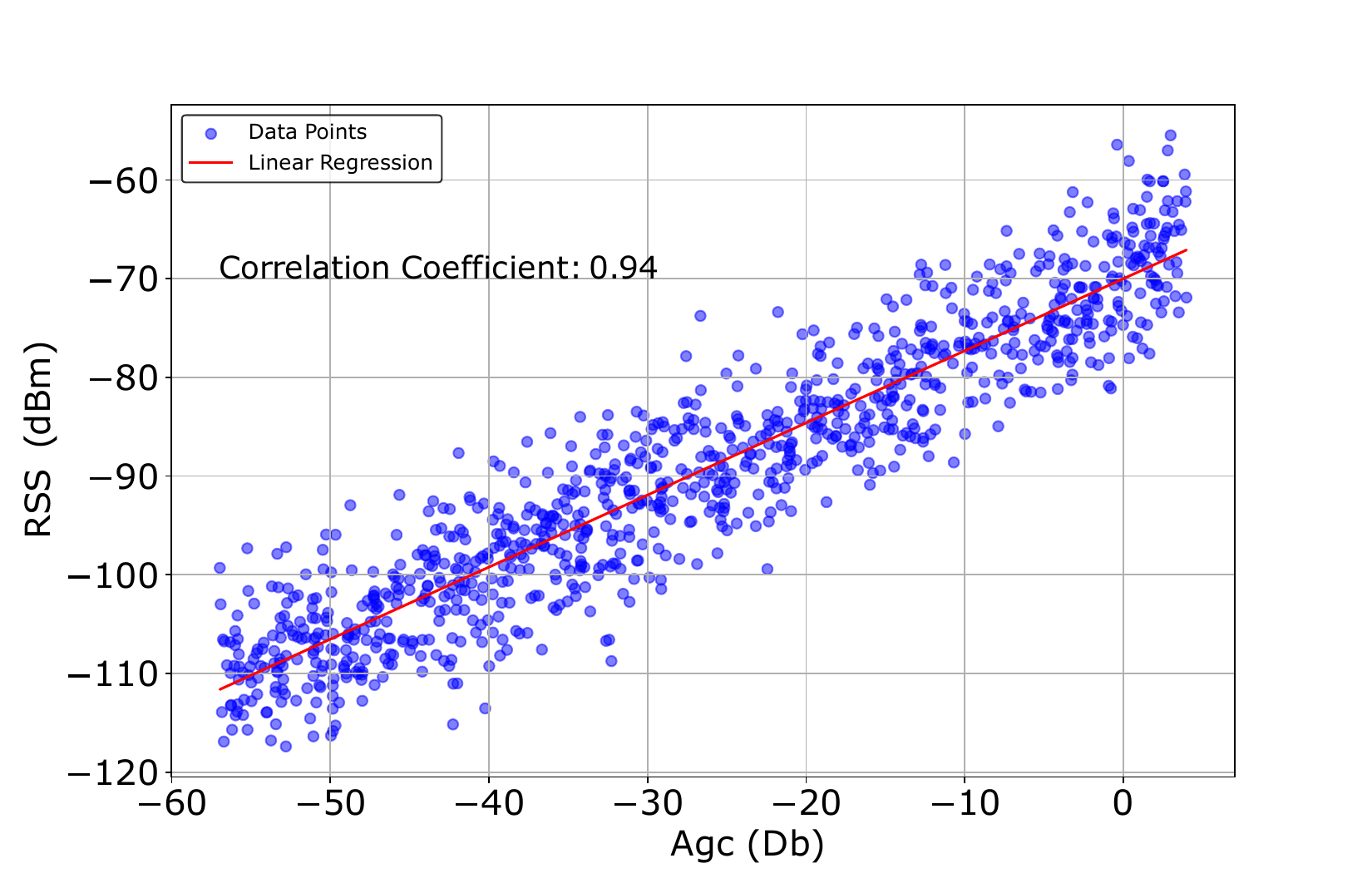}
    \caption{\small {Correlation between Agc and RSS}}
    \label{fig:rssicorr}
\end{center}
\end{figure}

\subsection{Correlation Results across the GPS Parameters: Feature Selection and LDA}

\begin{figure}[!htb]
    \centering
    \includegraphics[width=\columnwidth, trim={0cm 0cm 0cm 0.9cm}]{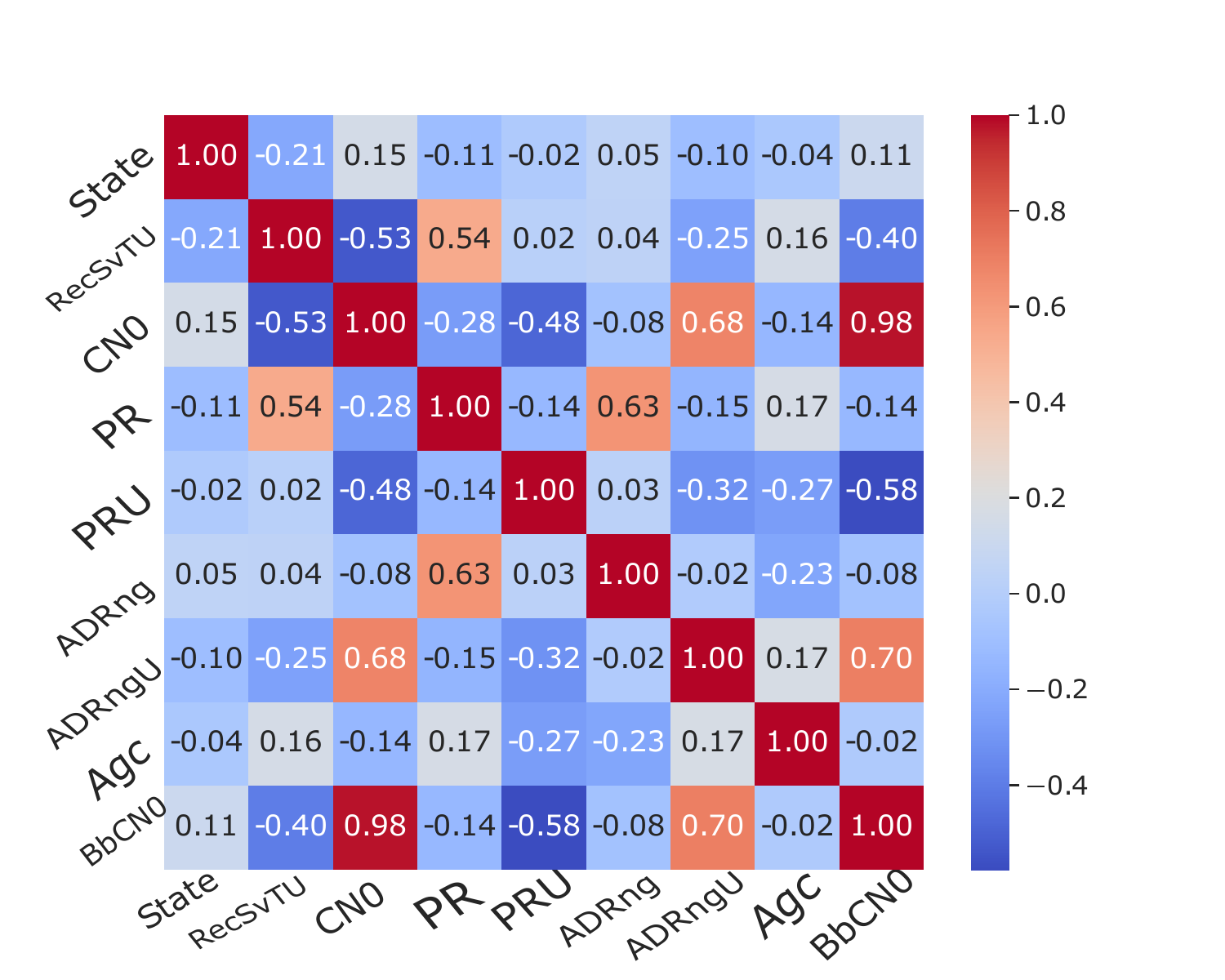} 
    \caption{\small {Correlation of various semi-processed GPS parameters}}
    \label{fig:corrheat}
\end{figure}

Next, we consider all our data: our traces and the Kaggle GNSS traces. The trends are roughly similar. 
We plot the correlation for all pairs of GPS signals (results shown in Figure~\ref{fig:corrheat}) $\cno$ exhibits a
strong correlation with $\bbcno$ (0.98). Hence, only one of them is required. We choose $\cno$.
The correlation of $\cno$ with other signals is as follows: 
$\adru$ (0.68), $\pru$ (0.48), and $\rsvtu$ (-0.53).
$\pr$ shows a
correlation of 0.63 and 0.54 with $\adr$ and $\rsvtu$, respectively. This means that we have a lot of pairs of 
parameters that have reasonably high correlations. This motivates the use of a feature reduction algorithm
such as LDA.
\section{model architecture}
\label{model}
We shall use the same ML model to classify both the ambience as well as human
activity. Henceforth, we will refer to
both an ambient setting as well as human activity as an \textit{event}. 
Figure~\ref{fig:mlflowdiagram} shows our proposed ambient
sensing framework. 

\begin{figure}[!htb]
    \begin{center}
    \includegraphics[scale=0.28]{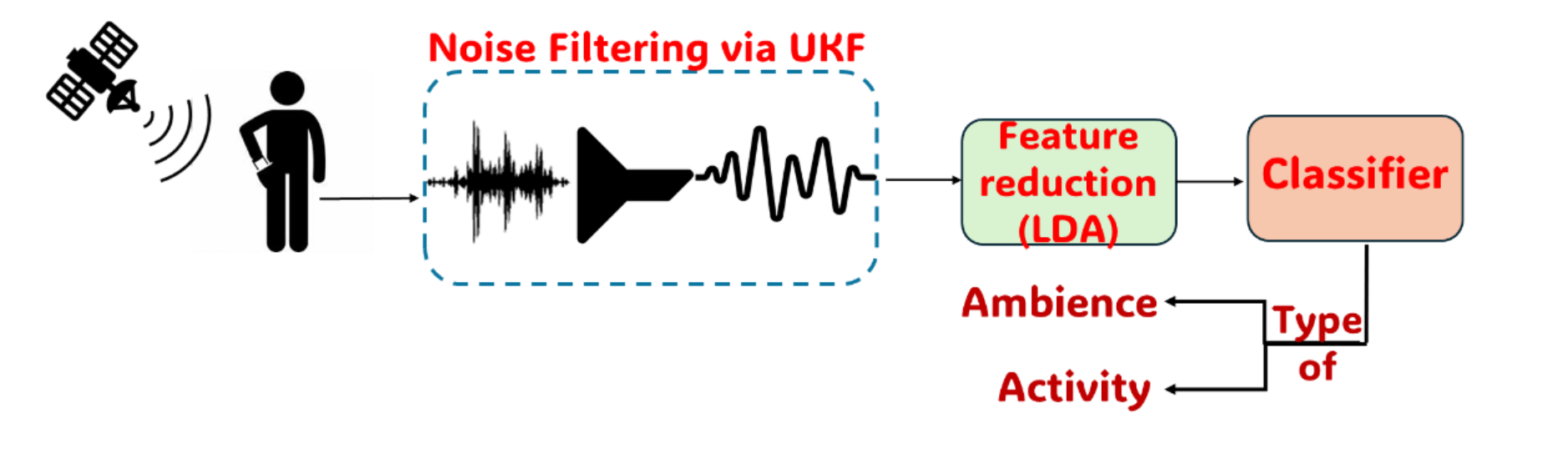}
                \caption{Flow diagram of our ML model}
    \label{fig:mlflowdiagram}
    \end{center}
\end{figure}

We first collect 8 features from the phone. Recall that we had discarded $\bbcno$. 
Then, we eliminate the nonlinear noise using Unscented Kalman filtering (UKF).
Subsequently, we perform feature reduction using the LDA algorithm. We need to classify the features
into $C$ classes, then the output is a vector with $C-1$ elements (dimensions reduced).
Then we use an ML-based classifer to map the reduced set of features to a class label. In
the evaluation section (Section~\ref{eval}), we shall evaluate different types of ML models. 

\subsection{{Preprocessing}}
\label{preprocess}
\paragraph{\textbf{\textit{Noise Filtering}}} GPS signals are susceptible to multipath effects, resulting in noise that affects the accuracy of activity or ambient-recognition models. 
While the Kalman Filter effectively filters noise in linear systems, it is inadequate for the nonlinear dynamics inherent in GPS signals~\cite{julier1997new}. We thus
use the Unscented Kalman Filter (UKF)~\cite{hu2020unscented}, which is
adept at handling these nonlinearities by employing a deterministic sampling approach.
By selectively filtering the \textit{non-critical} noise components, the UKF maintains a balance between noise
reduction and the retention of essential signal features. 

\subsection{Feature Reduction}
Prior to applying LDA, StandardScaler~\cite{ahsan2021effect} normalization was used to address the issue of 
differing scales and
normalize the semi-processed parameters. The output is a $C-1$ element vector. 
\section{Ambient Sensing and Human Activity Recognition}
We group ambient sensing and human activity recognition (HAR) into one section because the techniques that are
used are similar. 
\label{eval}

\subsection{Evaluation Setup}
\subsubsection{ML Models}
We evaluated six machine
learning algorithms: Random Forest (RF), K-Nearest Neighbor (KNN), Support Vector Machine (SVM), Decision Tree (DT),
Naive Bayes (NB) and Gradient Boosting (GB). Given that we were getting a high accuracy, we did not use
more sophisticated CNN-based models. These algorithms were chosen due to their popularity in the related
work~\cite{guo2019wiar, vidya2022wearable}. Some of them also produce explainable models and results.
To ensure the model's generalizability to new data and avoid overfitting,
extensive empirical analyses were conducted to determine the optimal hyperparameters for each model using 
GridSearchCV~\cite{kong2021exploratory} 
(refer to Table \ref{tab:hyper}). 

\begin{table}[ht]
\centering
\caption{Hyperparameters used for each classifier \label{tab:hyper}}
\begin{tabular}{lp{0.25\textwidth}}
\toprule
\textbf{Classifier} & \textbf{Hyperparameters} \\ 
\midrule
Random Forest & n\_estimators: 100, max\_depth: 10, min\_samples\_split: 10, min\_samples\_leaf: 4, bootstrap: True \\ 
\midrule
Support Vector Machine & C: 110, kernel: rbf, tol: 0.001, break\_ties: True \\ 
\midrule
K-Nearest Neighbors & n\_neighbors: 10, weights: distance, algorithm: auto, leaf\_size: 30 \\ 
\midrule
Decision Tree & max\_depth: 20, min\_samples\_split: 20 \\ 
\midrule
Naive Bayes & var\_smoothing: 1e-9, fit\_prior: True, class\_prior: None \\ 
\midrule
Gradient Boosting Classifier & n\_estimators: 100, learning\_rate: 0.01, max\_depth: 10, min\_samples\_split: 10, min\_samples\_leaf: 4 \\ 
\bottomrule
\end{tabular}
\end{table}

Notably, for SVM, we adopted the one-versus-one (OVO) strategy to address class imbalance, transforming the multi-class
classification problem into multiple binary classification tasks~\cite{daengduang2016enhancing}. Additionally, a
ten-fold cross-validation method was used to minimize overfitting. The dataset was randomly partitioned into 80\% for
training and 20\% for testing.

\subsubsection{Details of the Platform} The learning models were trained on the Google
Collaboratory cloud platform\footnote[1]{https://colab.research.google.com/}.
The cloud-based system was equipped
with a dual-core Intel\textsuperscript{\textregistered} Xeon\textsuperscript{\textregistered} CPU running at 2.20GHz,
13GB of RAM, and a disk capacity of 107.72 GB. The hyperparameter tuning task was slow and time consuming.
Hence, we used a
NVIDIA\textsuperscript{\textregistered} Tesla\textsuperscript{\textregistered} T4 GPU equipped with 15GB of dedicated
RAM.

\subsection{Ambient Sensing}
\subsubsection{Data collection}
For each ambient setting, over 100K samples were collected in various locations such as a dormitory floor, stadium,
bustling market and underground metro tunnel. The data collection involved 20 volunteers 
who were
graduate and undergraduate students in a premier university. 
The were aged between 19 and 28. The volunteers were first rigorously trained on how to take
measurements. Furthermore, to ensure the generality of the dataset, the volunteers were
requested to vary the settings within limits. For example, one group put the mobile phones in their
shirt pockets, some kept it in their pant pockets and some in their purses. 
 
\subsubsection{Class Labels}
Five user environments were defined: flight\footnote{All the signals were collected in passive, receive-only mode. There was no signal transmission. This is allowed as per rules.}, indoor, metro tunnel, open ground and outdoor crowded area. Indoors, open
areas and outdoor crowded areas are distinguishable using precise location coordinates and mapping services like
Google Earth\footnote[2]{https://earth.google.com/web/}. This is because multiple locations belonging to different classes
maybe in close proximity and it proved to be hard to distinguish them in our experiments. In any case, the objective of this work is to use
GPS semi-processed data alone.
For the open space and
outdoor crowded area  settings, samples
were collected in an open-air stadium. 
We considered two cases: empty and filled (with people). 

\subsubsection{Results with Parameters Considered Individually}
We evaluated the effectiveness of each of the semi-processed GPS parameters (after Kalman Filtering (UKF)) for
characterizing the ambient environment 
(see Figure~\ref{fig:ambientperparam}). 

\begin{figure*}[htb]
    \begin{center}
    \includegraphics[width=0.8\textwidth]{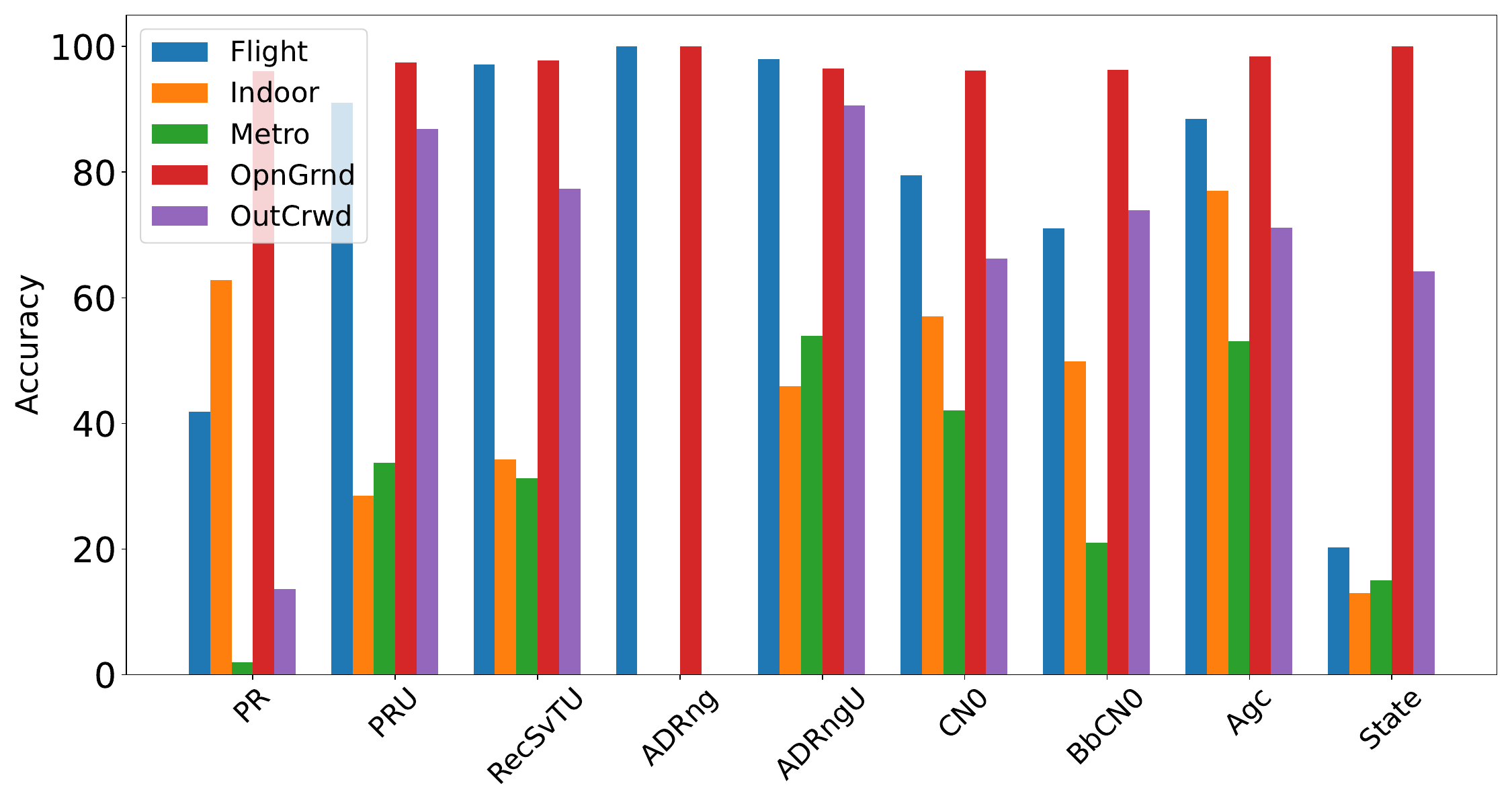}
            \caption{Ambient sensing accuracy for each semi-processed GPS parameter}
    \label{fig:ambientperparam}
    \end{center}
\end{figure*}

In the interest of space, only the most accurate results for each parameter (across all models) have been shown. 
$\agc$ classifies
the ambient with an average accuracy of 86.7\% while $\adr$ performs the worst with an accuracy of 40.6\%. All other
parameters have an accuracy range between 57.9-82.1\%. This accuracy is on the lower side because we are
considering the parameters {\em individually}. We combine the parameters to achieve better results.

\subsubsection{Results with Fused Parameters}

\begin{figure*}[htb]
\centering
    \includegraphics[scale=0.55]{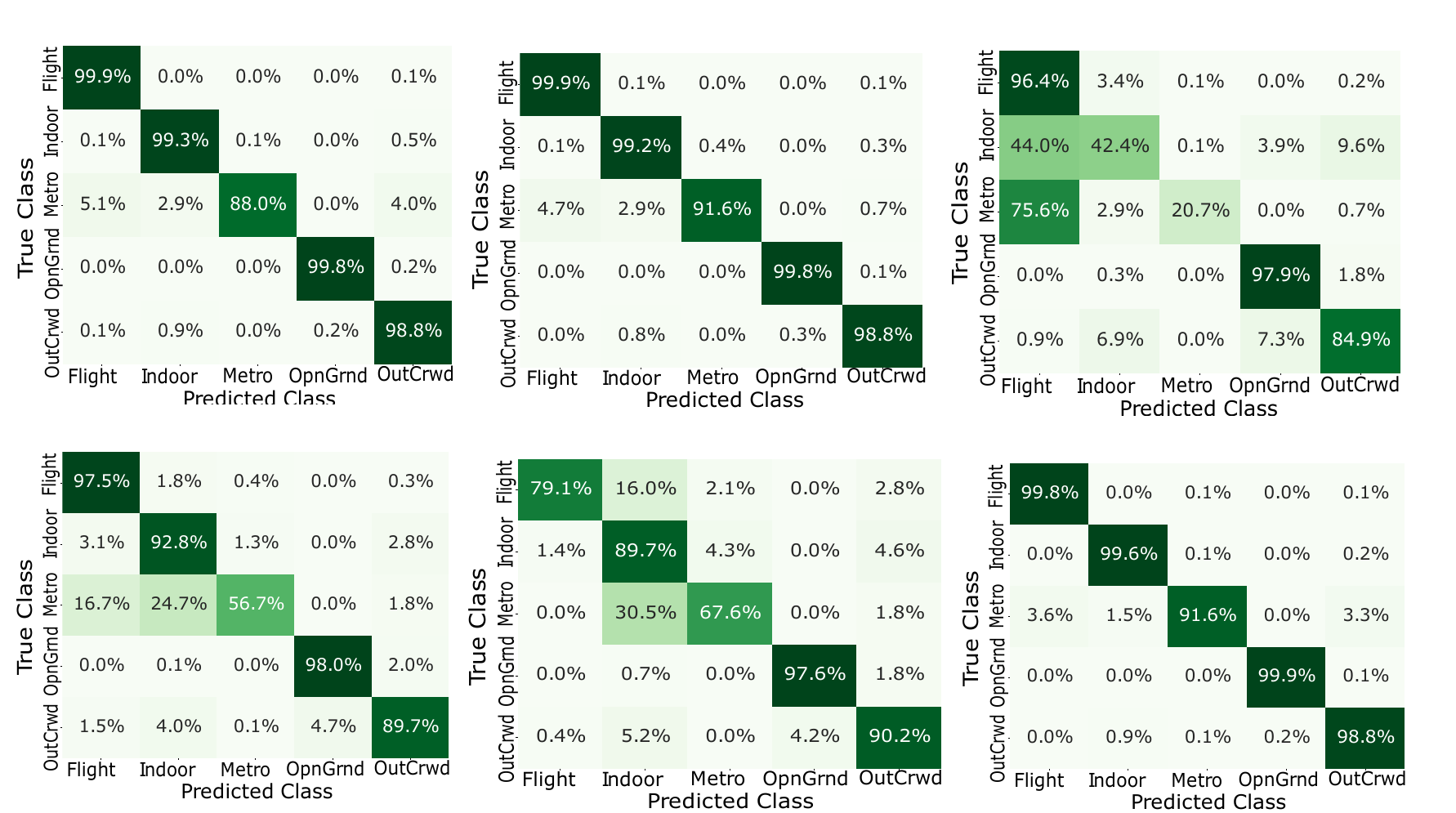}
    \caption{  { Ambient classification confusion matrix: {\bf(a)} RF, {\bf (b)} DT}, {\bf (c)} SVM, {\bf (d)} KNN, {\bf (e)} NB, and {\bf (f)} GB}
    \label{fig:ambientheatmap}
\end{figure*}

The accuracy of the models after all the steps (UKF + scaling + LDA) is shown in
Figure~\ref{fig:ambientheatmap}. 
Although SVM accurately predicted the results for three settings, it failed to classify the Indoor and Metro environments with
misclassification rates of 57.6\% and 79.3\%, respectively. Similarly, KNN achieved over 90\% accuracy for all the
settings other than Metro, where it had a misclassification rate of 43.4\%. 
The NB (Naive Bayes) model also underperformed for Metro with a
misclassification rate of 32.4\%. The Metro setting is hard to classify because several
parameters such as $\adr$ and $\state$ are not available. 
In contrast, RF (Random Forest) and GB (Gradient Boosting) exhibited the best performance with accuracies
exceeding 90\% for all the settings.

\begin{table}[!htb]
\small{ 
\centering{%
\caption{Performance metrics of ambient classifier models: GB and RF are the best. The results are averaged across
all the phones and volunteers.}
\label{tab:performance-metricsambience}
\begin{tabular}{|l|l|l|l|l|l|}
\hline
\textbf{Classifier} & \textbf{Ambience} & \textbf{Acc} & \textbf{Sen} & \textbf{Spe} & \textbf{F-score} \\ \hline
GB & Flight & 99.8 & 99.8 & 99.9 & 99.8 \\ \cline{2-6} 
 & Indoor & 99.6 & 99.6 & 99.8 & 99.4 \\ \cline{2-6} 
 & Metro & 91.6 & 91.6 & 99.9 & 93.9 \\ \cline{2-6} 
 & OpnGrnd & 99.9 & 99.9 & 100.0 & 99.9 \\ \cline{2-6} 
 & OutCrwd & 98.8 & 98.8 & 99.8 & 98.7 \\  \hline
RF & Flight & 99.5 & 99.9 & 99.9 & 99.8 \\ \cline{2-6} 
 & Indoor & 99.3 & 99.3 & 99.8 & 99.2 \\ \cline{2-6} 
 & Metro & 88 & 88.0 & 100.0 & 93.3 \\ \cline{2-6} 
 & OpnGrnd & 99.8 & 99.8 & 100.0 & 99.9 \\ \cline{2-6} 
 & OutCrwd & 98.8 & 98.8 & 99.7 & 98.4 \\  \hline
DT & Flight & 99.9 & 99.9 & 99.9 & 99.8 \\ \cline{2-6} 
 & Indoor & 99.2 & 99.2 & 99.8 & 99.1 \\ \cline{2-6} 
 & Metro & 91.6 & 91.6 & 99.9 & 93.2 \\ \cline{2-6} 
 & OpnGrnd & 99.8 & 99.8 & 99.9 & 99.9 \\ \cline{2-6} 
 & OutCrwd & 98.8 & 98.8 & 99.9 & 98.9 \\  \hline
KNN & Flight & 97.5 & 97.5 & 98.7 & 97.2 \\ \cline{2-6} 
 & Indoor & 92.8 & 92.8 & 98.4 & 92.4 \\ \cline{2-6} 
 & Metro & 56.7 & 56.7 & 99.7 & 62.2 \\ \cline{2-6} 
 & OpnGrnd & 98 & 98.0 & 99.0 & 98.3 \\ \cline{2-6} 
 & OutCrwd & 89.7 & 89.7 & 98.4 & 89.1 \\  \hline
NB & Flight & 79.1 & 79.1 & 99.6 & 87.8 \\ \cline{2-6} 
 & Indoor & 89.7 & 89.7 & 93.0 & 79.6 \\ \cline{2-6} 
 & Metro & 67.6 & 67.6 & 98.7 & 51.0 \\ \cline{2-6} 
 & OpnGrnd & 97.6 & 97.6 & 99.1 & 98.2 \\ \cline{2-6} 
 & OutCrwd & 90.2 & 90.2 & 97.4 & 86.0 \\  \hline
 SVM & Flight & 96.4 & 96.4 & 88.4 & 85.4 \\ \cline{2-6} 
 & Indoor & 42.4 & 42.4 & 97.7 & 55.0 \\ \cline{2-6} 
 & Metro & 20.7 & 20.7 & 100.0 & 33.7 \\ \cline{2-6} 
 & OpnGrnd & 97.9 & 97.9 & 97.4 & 97.2 \\ \cline{2-6} 
 & OutCrwd & 84.9 & 84.9 & 97.3 & 82.9 \\  \hline
\end{tabular}%
}}
\end{table}

Table~\ref{tab:performance-metricsambience} 
shows the same data for each setting using several popular ML metrics: accuracy, sensitivity,
specificity and the F-score. We find RF and GB to be the best models across metrics.

\subsubsection{Characterization of the Indoor Environment}

\begin{figure}[!htb]
    \centering
    \includegraphics[width=0.65\columnwidth]{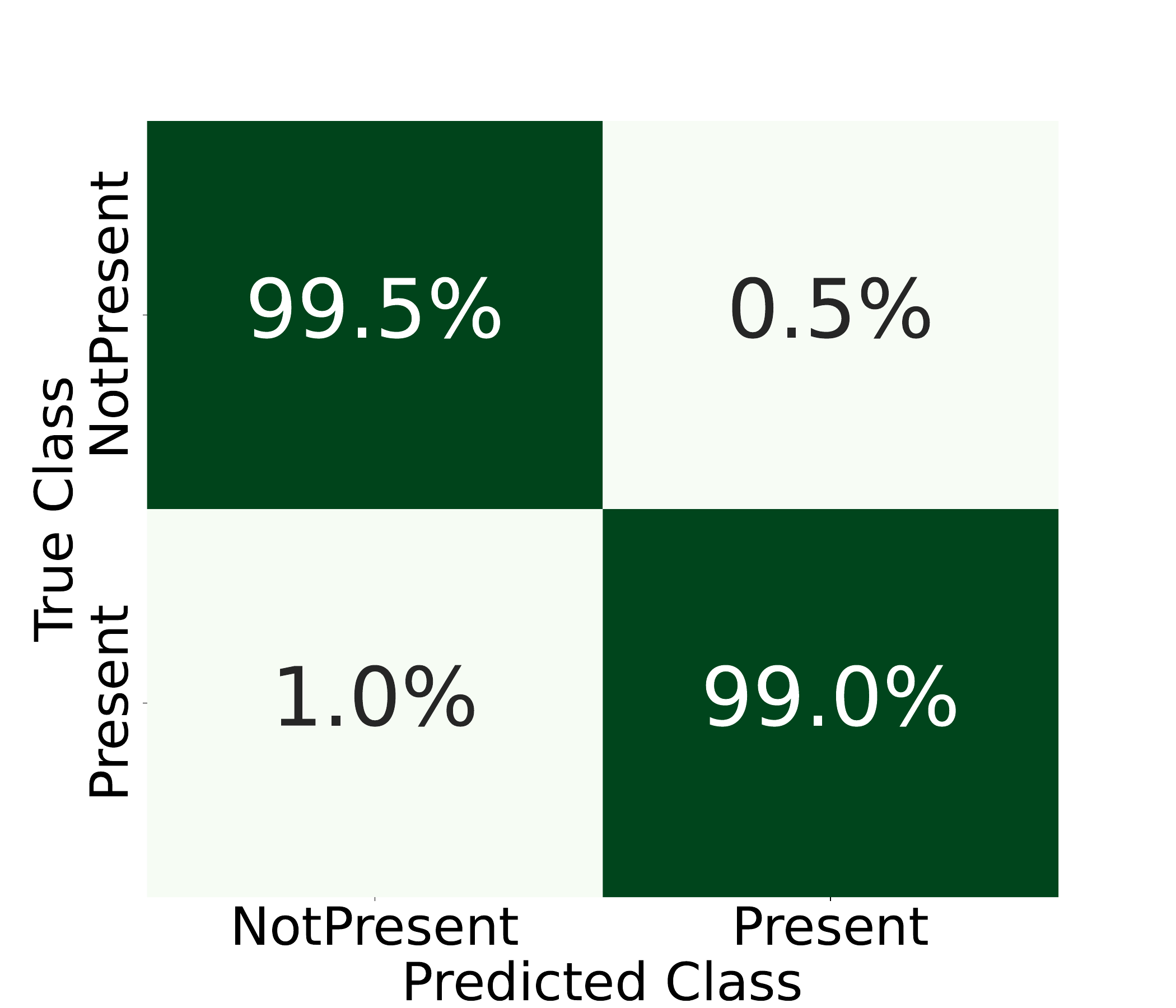} 
    \caption{\small {Confusion matrix of indoor environment with different population density}}
    \label{fig:presentnotpresent}
\end{figure}

Additionally, we further characterized the indoor environment based on population density. 
We consider two settings: nobody is present in a small (12 ft x 8 ft) room and two people are
present. The GB model accurately classified both scenarios with an accuracy of 99.25\%, as
shown in Figure~\ref{fig:presentnotpresent}.

\subsubsection{Relative Importance of Features}

\begin{figure}[htb]
\centering
          \includegraphics[scale=0.15]{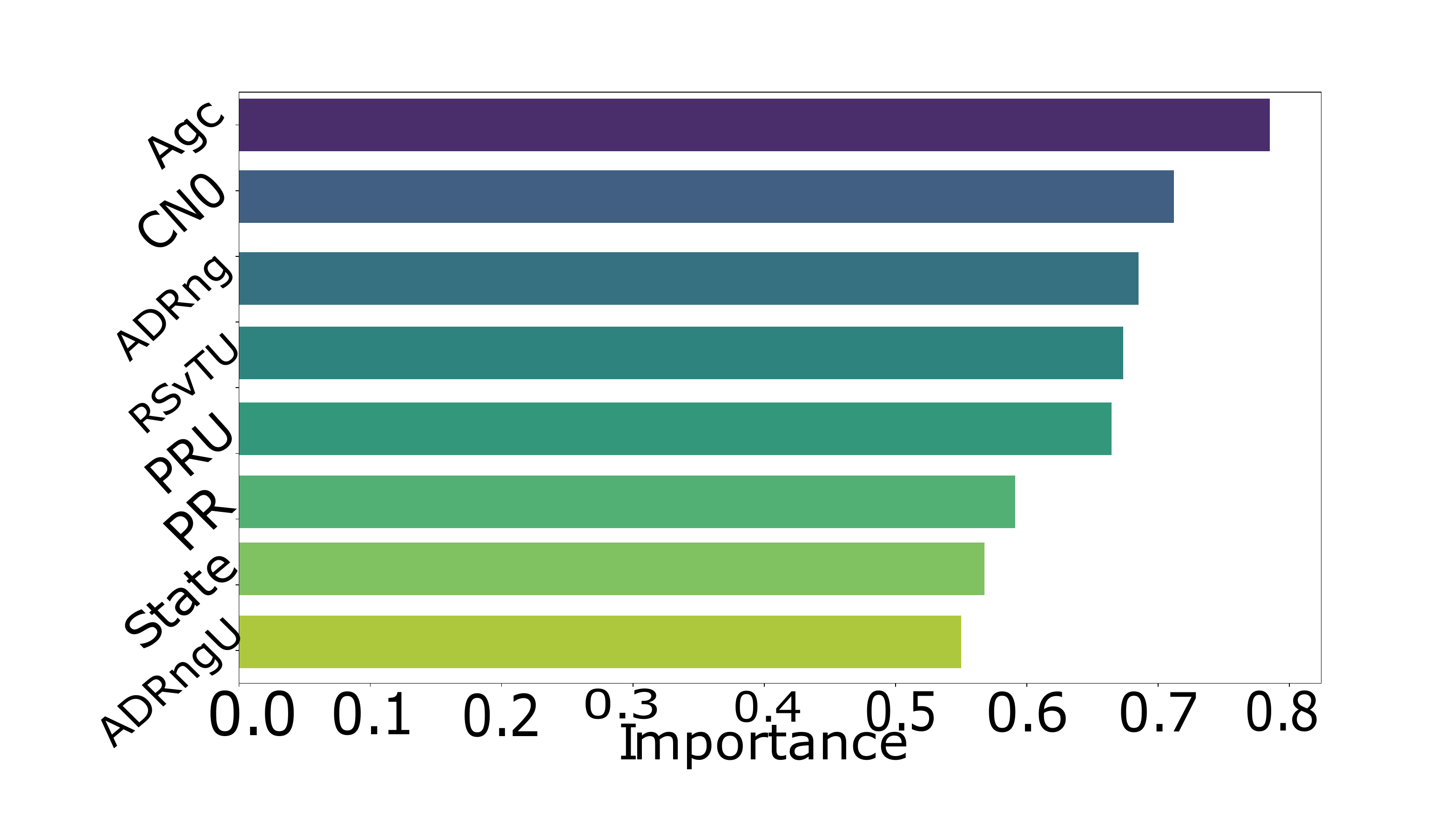}
  \caption{\small Feature importance test for ambient classification (GB model)}
  \label{fig:feaimpamb}
\end{figure}

We further evaluated the importance of each feature. 
This involves randomly shuffling the values of each parameter and measuring the resulting increase in the model's prediction error~\cite{featimp}. A feature is deemed important if shuffling its values leads to a significant increase in model error, indicating that the model relied on that feature for accurate predictions.
$\agc$ has the highest importance
score of 0.78, while all other features have importance scores ranging from 0.55 to 0.75 (see
Figure~\ref{fig:feaimpamb}), demonstrating their substantial contributions to classification.

\subsection{Human Activity Recognition(HAR)}

\subsubsection{Class Labels}
Prior work~\cite{ahmed2020enhanced, zheng2019zero, sekiguchi2021phased, gu2015paws, nafea2021sensor} has
extensively explored various HAR systems.
They cover both static activities (sitting, lying and standing) and dynamic
activities (such as walking, running or traveling in a vehicle).
The latter can be discerned on the basis of the speed or velocity.  The Android
platform's {\small $\mathtt{getSpeed}$} API~\cite{locapi} in the {\small $\mathtt{android.location.Location}$} package
facilitates this. In our case, we shall use
Doppler shifts~\cite{locapi} for this purpose (captured by the $\pr$ attribute).

Distinguishing static activities cannot be done on the basis of Doppler shifts.
We wish to classify static activities into four classes: sitting, standing, lying down and hand waving
(mobile phone not held with the moving hand). In case of hand waving, the
device was mounted on a wall stand and hand movements were performed in its close proximity (within 2m).

\subsubsection{Results with Individual GPS Parameters}

\begin{figure*}
    \begin{center}
    \includegraphics[width=0.8\textwidth]{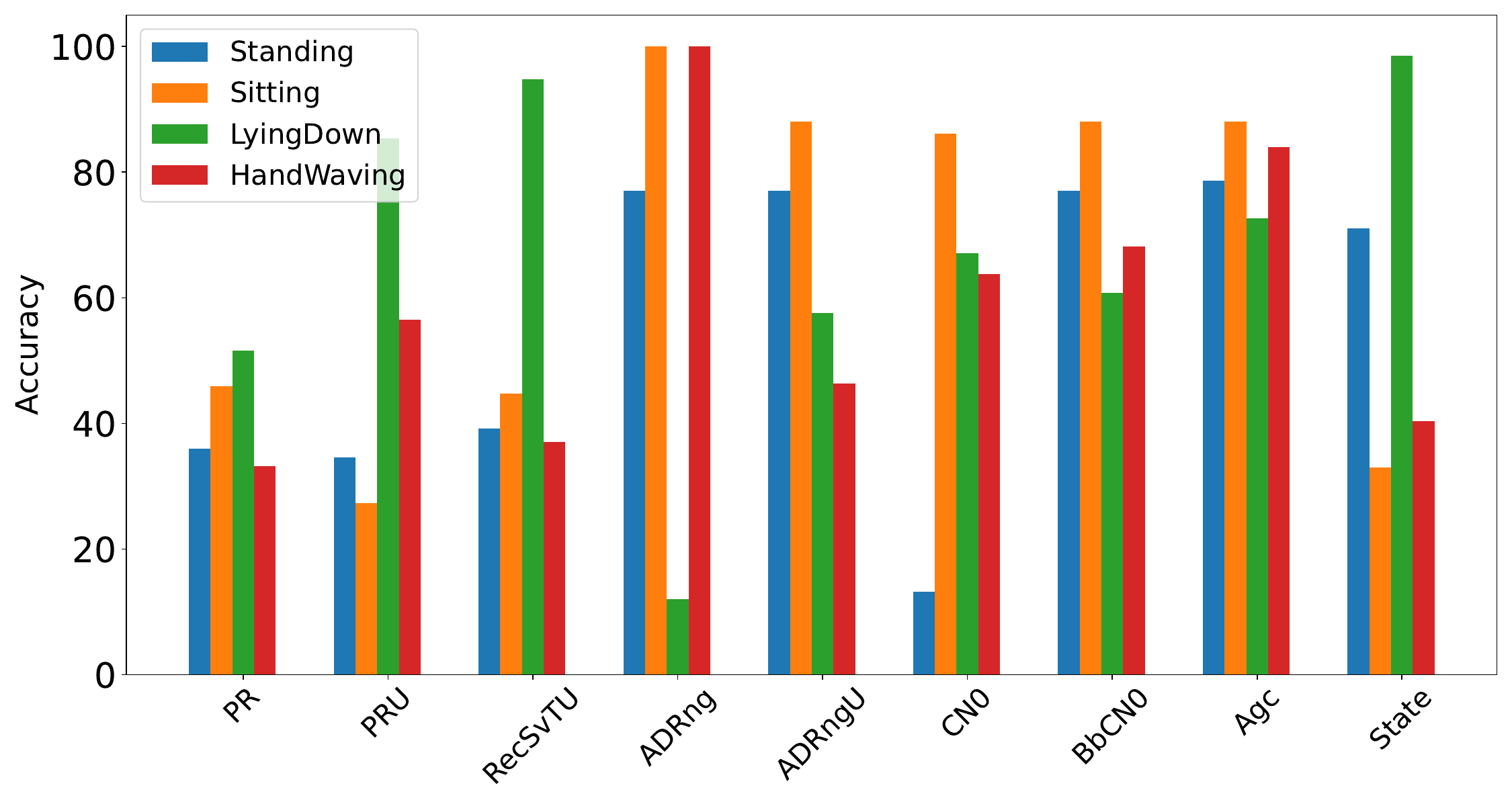}
        \caption{Activity classification accuracy for each semi-processed GPS parameter}
    \label{fig:activityperparam}
    \end{center}
\end{figure*}

Figure~\ref{fig:activityperparam} 
shows the accuracy of each activity for each semi-processed
GPS parameter. The results are consistent with previous observations. $\agc$ achieves the
highest accuracy of 77.2\%, while $\adr$ performs the
worst with a high misclassification rate of 72.6\%. The accuracy for
other parameters ranges from 35\% to 56\% reflecting the inability of individual parameters in classifying activities. Hence, we combine the parameters to achieve better results.

\subsubsection{Results with Fused Parameters}

\begin{figure*}[htb]
    \centering
    \includegraphics[scale=0.55]{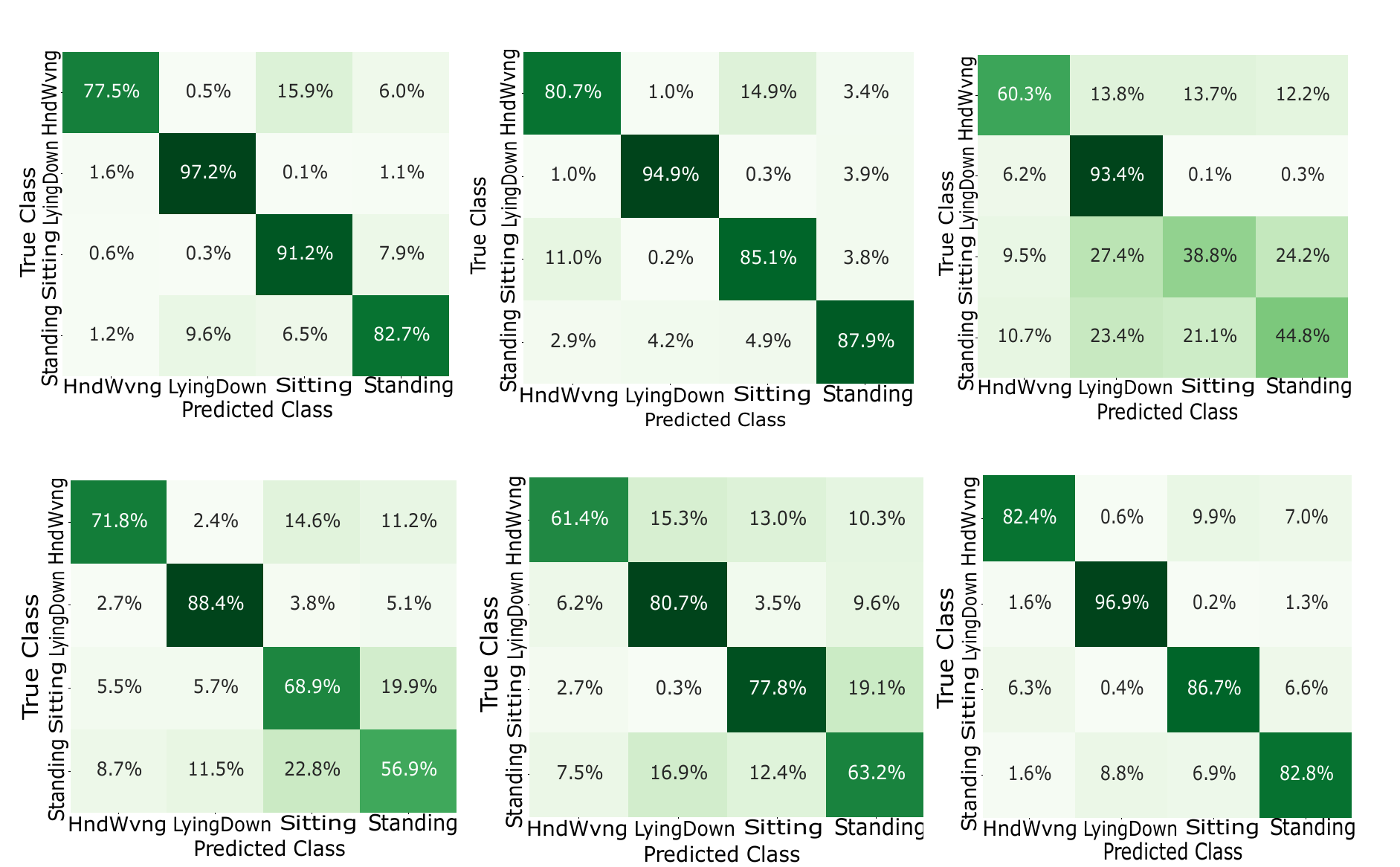}
    \caption{  { Activity classification confusion matrix: {\bf(a)} RF, {\bf (b)} DT}, {\bf (c)} SVM, {\bf (d)} KNN, {\bf (e)} NB, and {\bf (f)} GB}
    \label{fig:Activitytheatmap}
\end{figure*}

Figure~\ref{fig:Activitytheatmap}
compares model performance using the fused parameters. 
SVM achieves 93.4\% accuracy for lying down but has a
an accuracy of 60\% for other activities. KNN accurately classifies hand waving with an accuracy of
88.4\% but has an
average accuracy of 74.4\% for other activities. 
NB identifies the lying down posture with $\geq 80\%$ accuracy but
performs poorly (about 67\% on average) for the other three activities. 
RF, DT, and GB are the best -- they achieve a minimum accuracy of
80\%. The peak accuracy is 97.2\% (RF).

\begin{table}[!htb]
\small{ 
\centering{
\caption{Performance metrics of activity classifier models}
\label{tab:performance-metricsactivity}
\begin{tabular}{|l|l|l|l|l|l|}
\hline
\textbf{Classifier} & \textbf{Activity} & \textbf{Acc} & \textbf{Sen} & \textbf{Spe} & \textbf{F-score} \\ \hline
RF & HandWaving & 77.5 & 77.5 & 98.9 & 85.6 \\ \cline{2-6} 
 & LyingDown & 97.2 & 97.2 & 96.5 & 92.6 \\ \cline{2-6} 
 & Sitting & 91.2 & 91.2 & 92.3 & 86.4 \\ \cline{2-6} 
 & Standing & 82.7 & 82.7 & 94.7 & 83.8 \\ \hline
DT & HandWaving & 80.7 & 80.7 & 94.7 & 81.7 \\ \cline{2-6} 
 & LyingDown & 84.9 & 94.9 & 98.2 & 94.2 \\ \cline{2-6} 
 & Sitting & 85.1 & 85.1 & 93.2 & 84.0 \\ \cline{2-6} 
 & Standing & 87.9 & 87.9 & 96.3 & 88.8 \\ \hline
 GB & HandWaving & 82.4 & 82.4 & 96.7 & 85.4 \\ \cline{2-6} 
 & LyingDown & 96.9 & 96.9 & 96.7 & 92.8 \\ \cline{2-6} 
 & Sitting & 86.7 & 86.7 & 94.1 & 85.9 \\ \cline{2-6} 
 & Standing & 82.8 & 82.8 & 94.8 & 84.0 \\ \hline
KNN & HandWaving & 71.8 & 71.8 & 94.2 & 75.4 \\ \cline{2-6} 
 & LyingDown & 88.4 & 88.4 & 93.4 & 83.3 \\ \cline{2-6} 
 & Sitting & 68.9 & 68.9 & 85.6 & 66.8 \\ \cline{2-6} 
 & Standing & 56.9 & 56.9 & 87.3 & 59.3 \\ \hline
NB & HandWaving & 61.4 & 61.4 & 94.6 & 68.8 \\ \cline{2-6} 
 & LyingDown & 80.7 & 80.7 & 89.5 & 73.8 \\ \cline{2-6} 
 & Sitting & 77.8 & 77.8 & 90.1 & 76.4 \\ \cline{2-6} 
 & Standing & 63.2 & 63.2 & 86.6 & 63.1 \\ \hline
 SVM & HandWaving & 60.3 & 60.3 & 91.0 & 63.8 \\ \cline{2-6} 
 & LyingDown & 93.4 & 93.4 & 78.1 & 68.7 \\ \cline{2-6} 
 & Sitting & 38.3 & 38.8 & 87.7 & 45.4 \\ \cline{2-6} 
 & Standing & 44.8 & 44.8 & 86.8 & 49.4 \\ \hline 
\end{tabular}%
}}
\end{table}

Table~\ref{tab:performance-metricsactivity}
shows the same data along with other ML metrics: sensitivity, specificity and the F-score.
We can observe that
SVM performs the worst (accuracy =  57.4\%) and RF performs the best (accuracy = 87\%).
Consequently, RF was selected as the model of choice for this activity. 

\subsubsection{Relative Importance of Features}

\begin{figure}[htb]
\centering
         \includegraphics[scale=0.15]{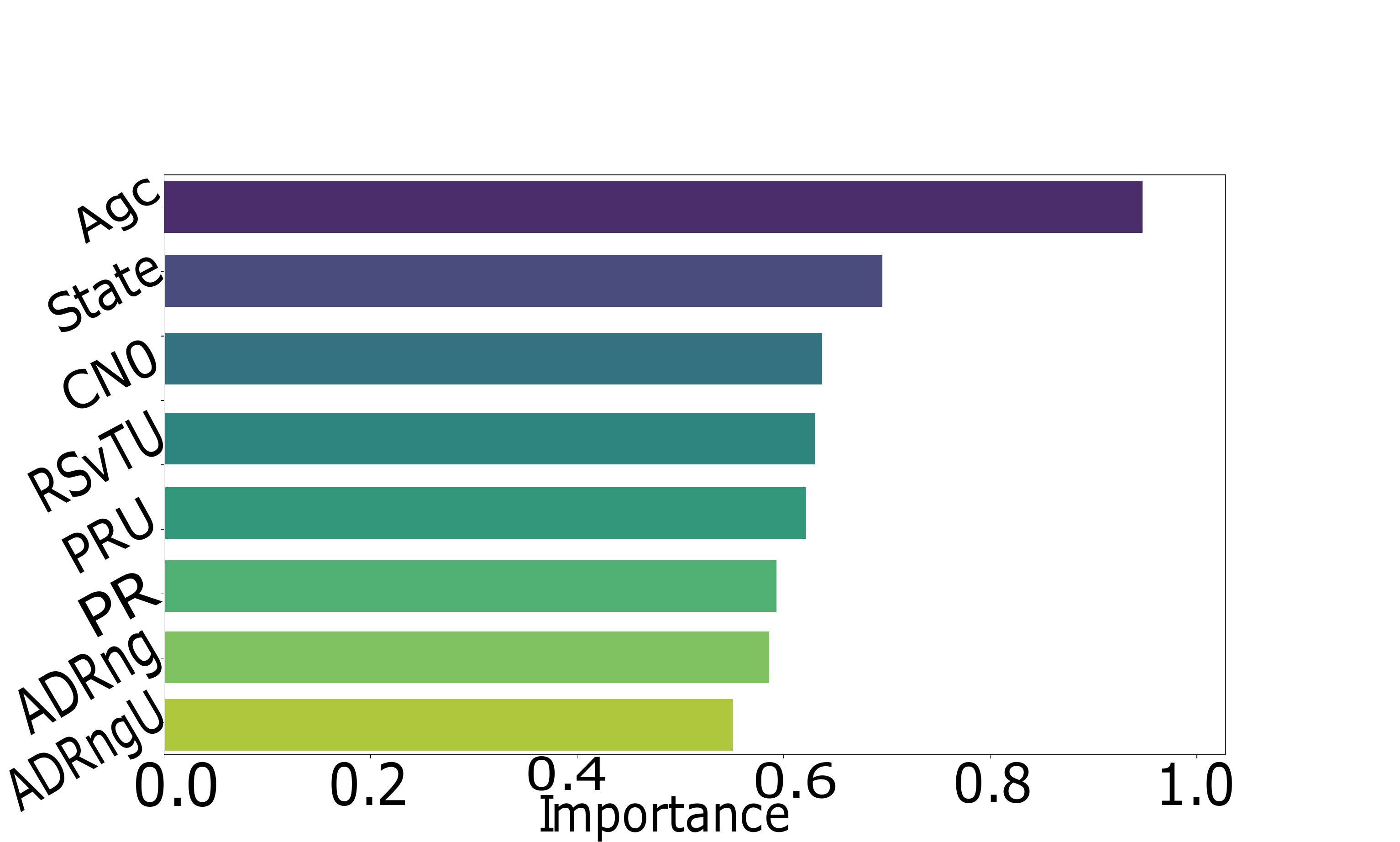}
  \caption{\small Feature importance test for human activity recognition}
  \label{fig:featimpact}
\end{figure}

Additionally, we evaluated the importance of each feature in HAR when they are fused together. The feature $\agc$ has the highest importance score of 0.88, while all other features have importance scores in the range of 0.5 to 0.65 (see Figure~\ref{fig:featimpact}). This further strengthens our claim -- using fused parameters improves accuracy.

\subsection{Assessment of the Robustness}
We conduct several experiments to demonstrate the robustness 
of our method. We consider different environmental conditions, variations in the satellite vehicle ID
(SvID) density, different test:train split, sparse fingerprinting techniques and accuracy variability across different phones.

\subsubsection{Variation in the Number of Unique SvIDs} 

\begin{figure}[!htbp]
\vspace*{0.5ex}
\vspace*{-1.5ex}
\hspace*{-2ex}
      \subfloat[]{%
          \includegraphics[scale=0.138]{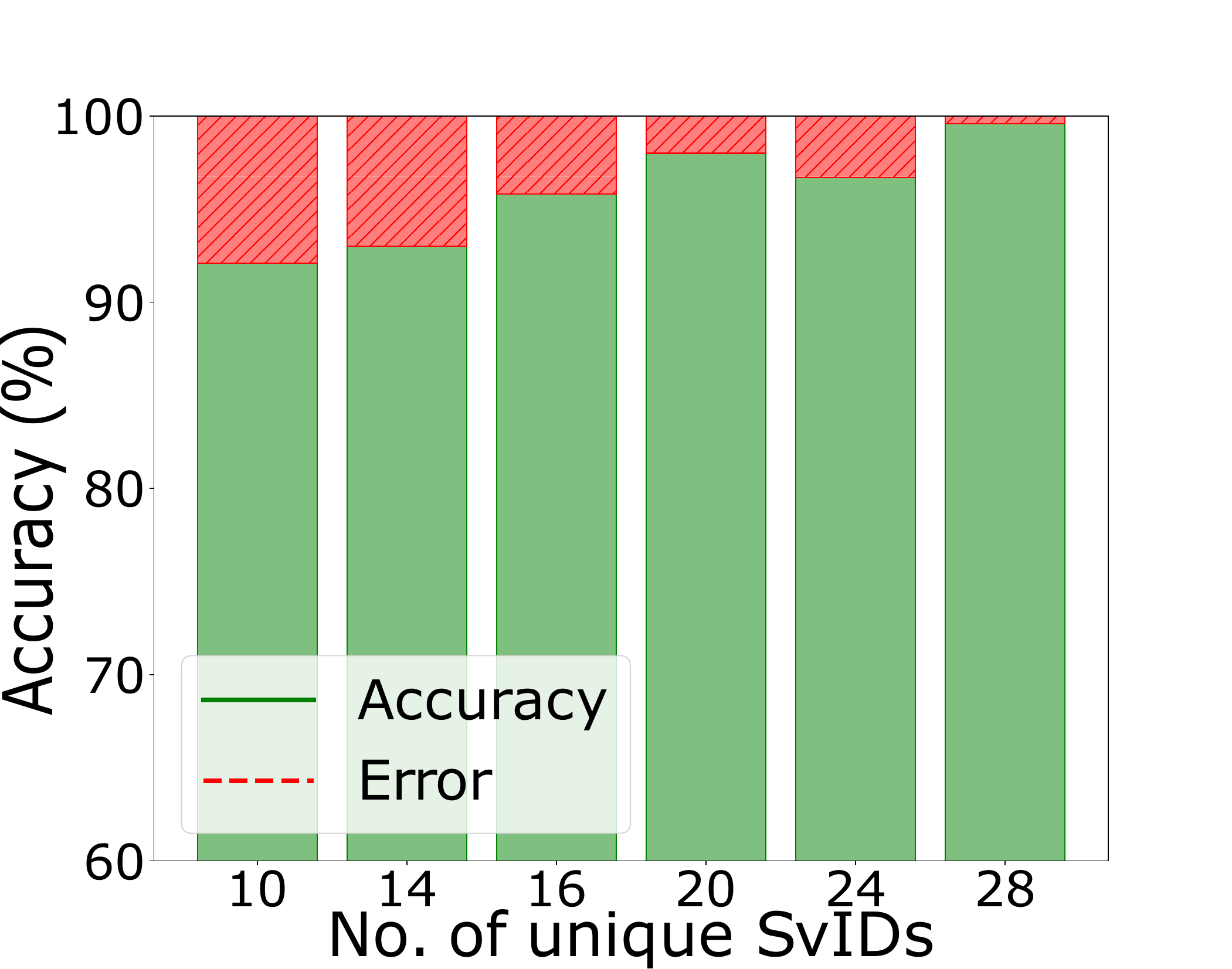}}
      \subfloat[]{%
\vspace*{-1.5ex}
\hspace*{-3ex}
         \includegraphics[scale=0.138]{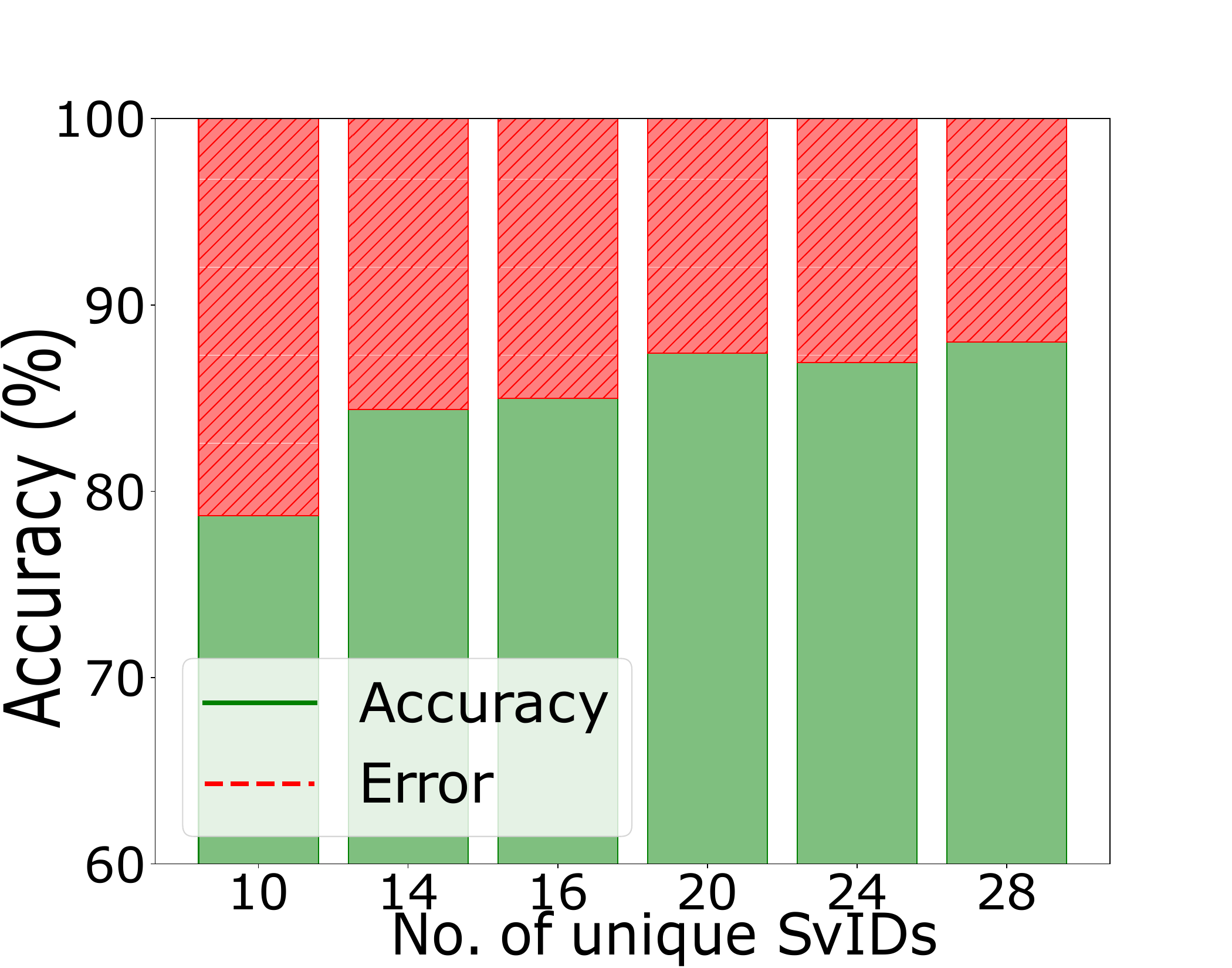}}
\vspace*{-1ex}
  \caption{\small  { Accuracy on varying the number of SvIDs for {\bf(a)} ambience classification and {\bf (b)} activity classification}}
  \label{fig:unseensvid}
\end{figure}

Figure~\ref{fig:unseensvid} 
shows the impact of varying the number of SvIDs on the prediction accuracy. A subset of SvIDs was randomly
selected from the logged data and only semi-processed GPS data corresponding to these SvIDs was used for evaluation.
Care was taken to maintain the integrity of the dataset by ensuring that SvIDs representing a significant portion of the
logged data were not removed. The dataset was becoming really sparse.

The maximum number of unique SvIDs observed during the experiment was 28. The results show that even with a SvID density
as low as 50\% of the total number of unique SvIDs (14 unique SvIDs) an accuracy of 92.1\% 
 (\ref{fig:unseensvid}(a)) and
84.4\%  (\ref{fig:unseensvid}(b)) in ambient and activity classification, respectively, was achieved. 

\subsubsection{Different Train:Test Splits}
Figure~\ref{fig:traintestratio} shows the accuracies when we vary the train:test ratio. For example, a ratio
of 80:20 indicates that the training data is 80\% of the overall dataset and the remaining 20\% of the data
is used for testing purposes. 

Even when the proportion of test data is as high as 50\%, the average accuracy of ambient classification is 94.2\% (\ref{fig:traintestratio}(a)), and 85.4\% (\ref{fig:traintestratio}(b)) for activity
classification, respectively. We can attribute this high degree of resilience to the UKF's noise removal and feature
preservation capabilities. When we remove UKF, the accuracy drops to roughly by more than 50\%.

\begin{figure}[!htbp]
\vspace*{0.5ex}
\vspace*{-1.5ex}
\hspace*{-2ex}
      \subfloat[]{%
          \includegraphics[scale=0.138]{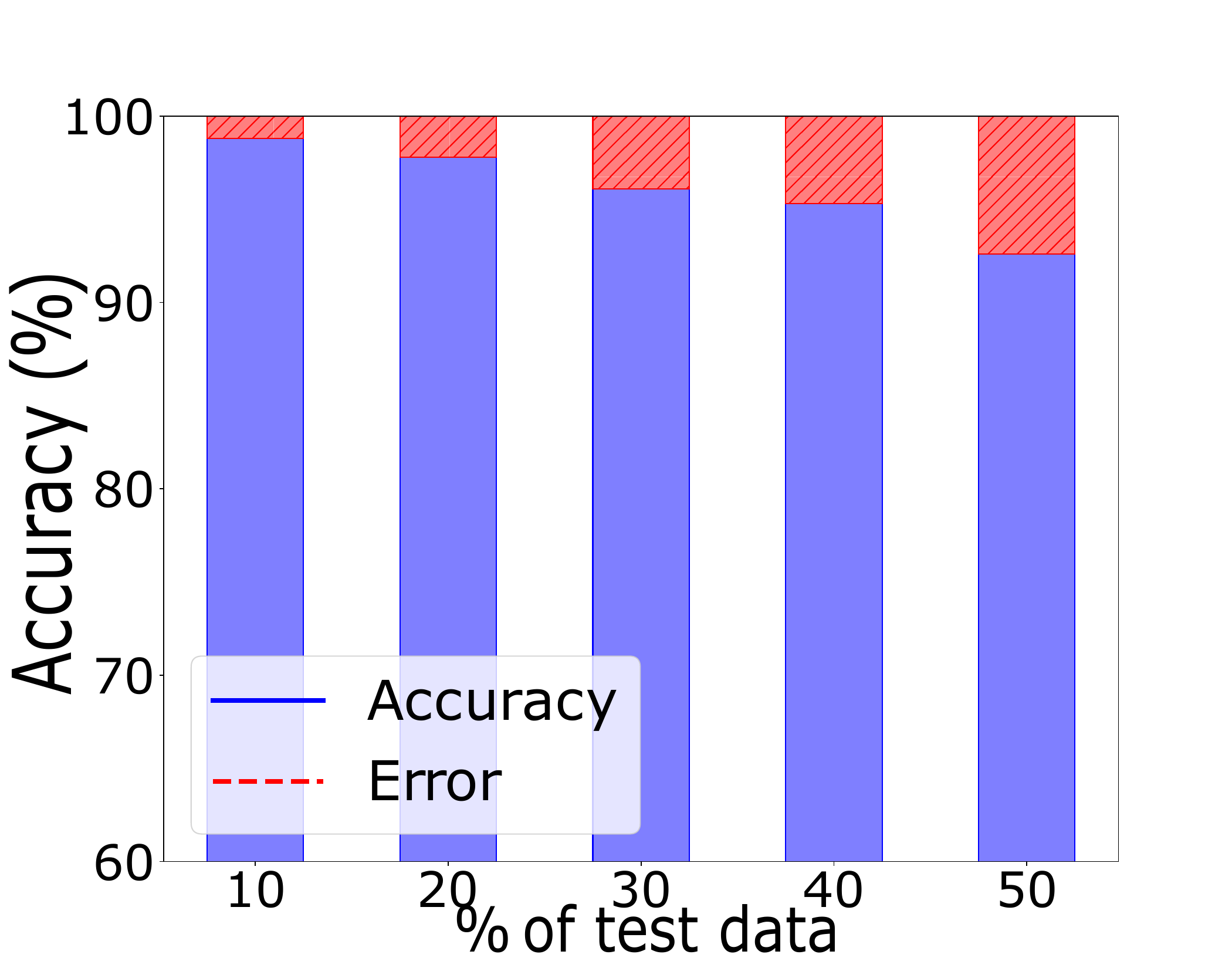}}
      \subfloat[]{%
\vspace*{-1.5ex}
\hspace*{-3ex}
         \includegraphics[scale=0.138]{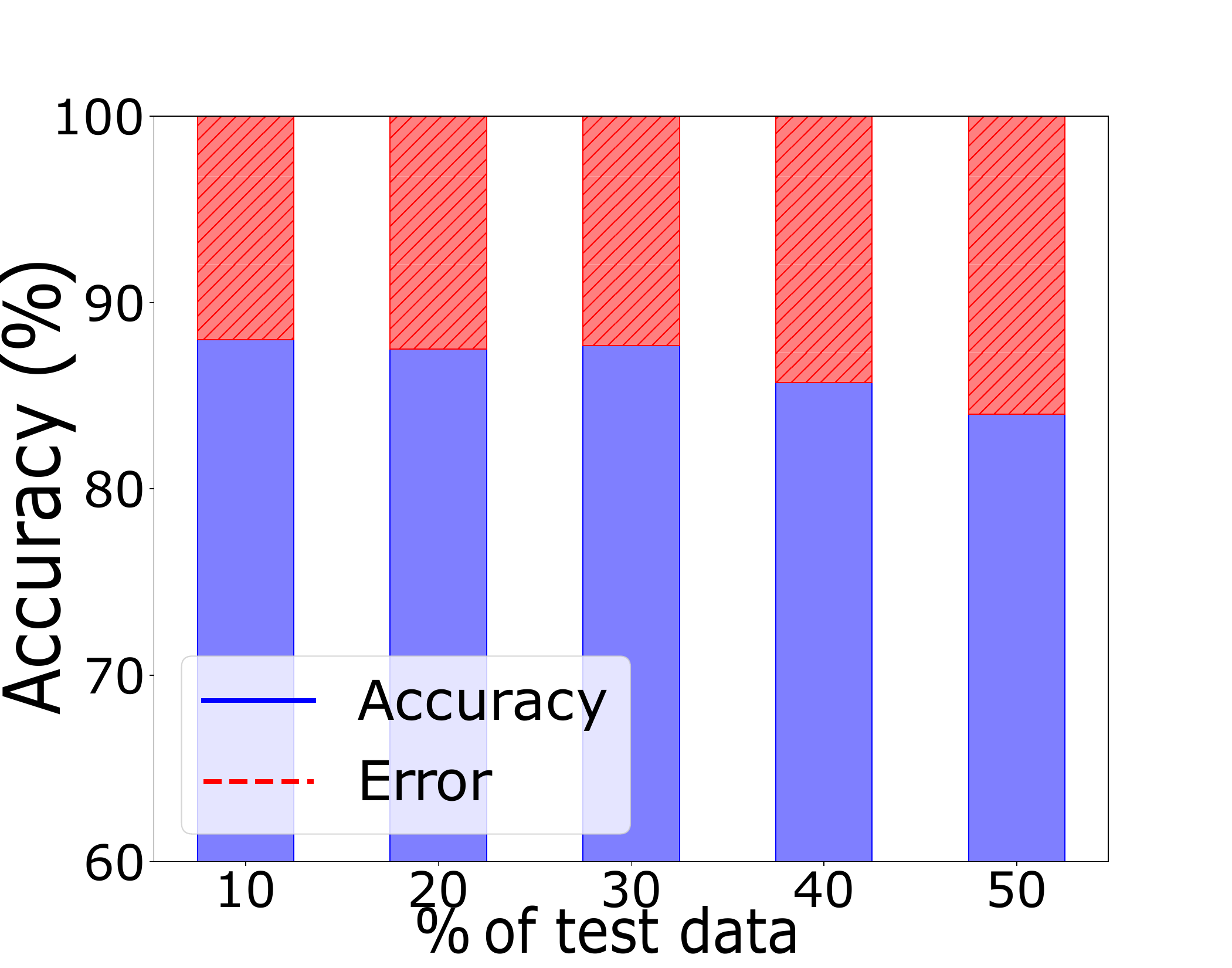}}
\vspace*{-1ex}
  \caption{\small  { Accuracy with different level of train:test ratio for {\bf(a)} ambience classification and {\bf (b)} activity classification}}
  \label{fig:traintestratio}
\vspace*{-3ex}
\end{figure}

\subsubsection{Unseen Events}
Additionally, the accuracy of the model was assessed by experiments performed by volunteers using another
set of Android phones on settings that are similar to the settings that we have been using up till now.
It is important to note that up till now we only used a set of 5 phones and the same settings even though
we varied the following parameters: time of the year, time of the day, weather condition and there was some
noise. For example, in a crowded area, the composition of the crowd varied. In this case, we vary 
everything other than the nature of the setting. For example, we collected measurements at locations
that were roughly 1000 kms away from our original location. One location had an altitude  of 3.5 kms.
We also tested on a cruise ship, which we categorized as an open space.

\begin{figure}[!htbp]
      \subfloat[]{%
          \includegraphics[scale=0.06]{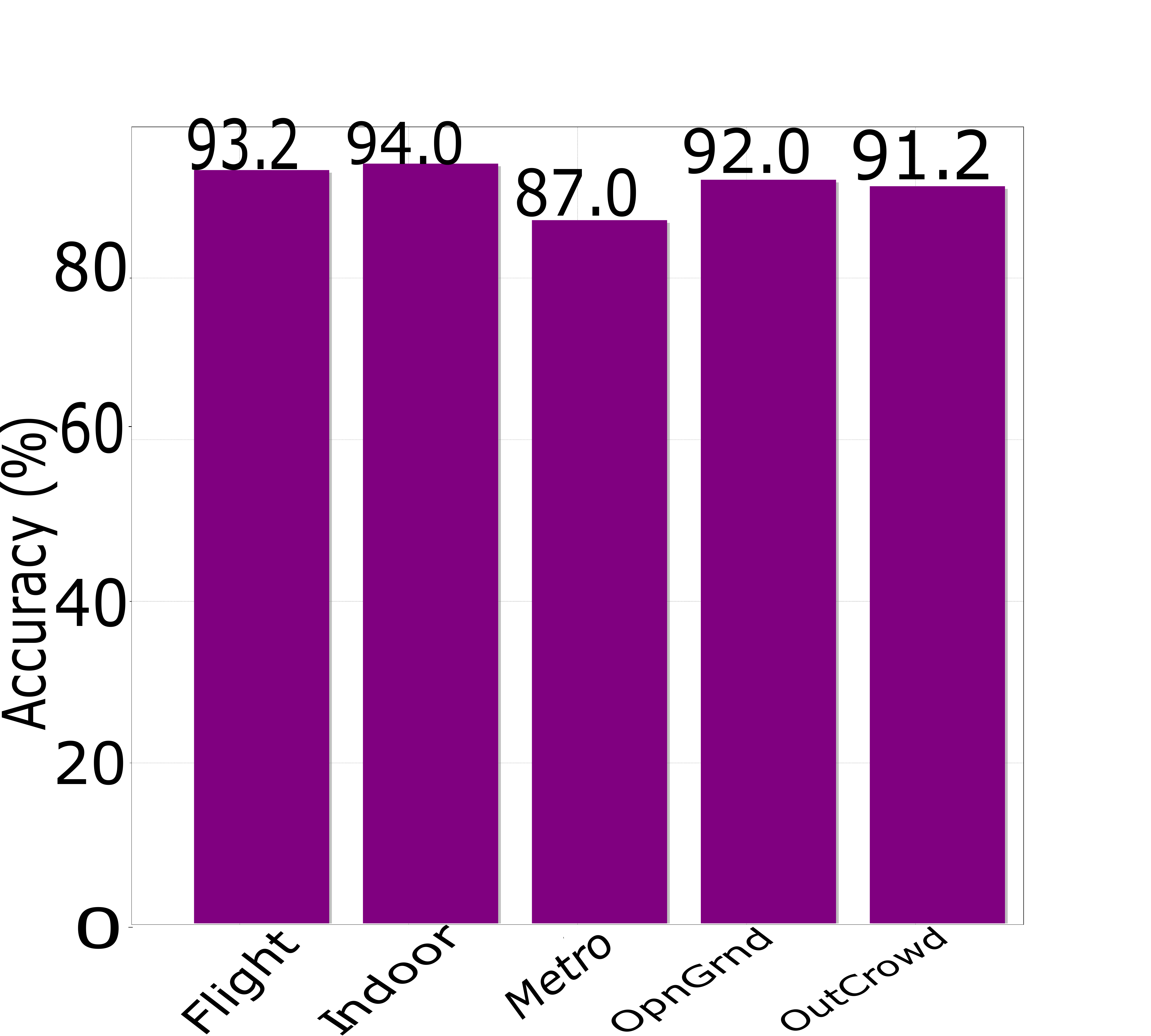}}
      \subfloat[]{%
         \includegraphics[scale=0.06]{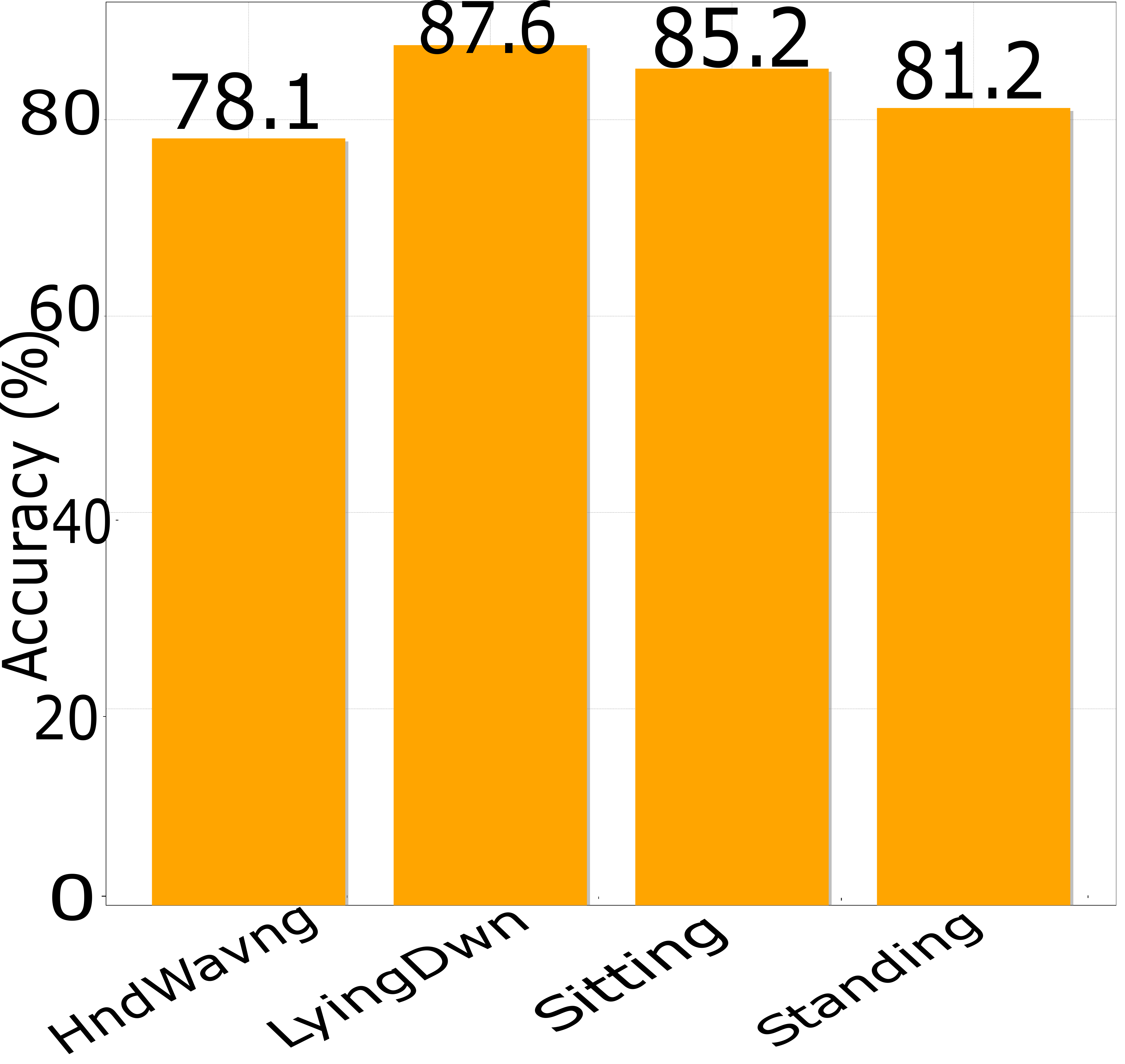}}

  \caption{\small  { Accuracy of events performed by random volunteers at random locations for {\bf(a)} ambience classification and {\bf (b)} activity classification}}
  \label{fig:unknownevent}
\end{figure}

Figure~\ref{fig:unknownevent}  shows the results.
While a decrease in accuracy was observed, it remained consistently above
91.5\%~\ref{fig:unknownevent}(a) for ambient recognition and 83\%~\ref{fig:unknownevent}(b) for activity classification,
respectively. The best-performing technique was still GB for ambient classification and RF for human activity recognition.
This reduction in precision was mainly due to variations in multipath effects and the significantly
different experimental setups. 
Moreover, these volunteers did not go through our orientation process.
There is a possibility that our orientation process may have sub-consciously created some
degree of homogeneity, even though we tried our best to avoid it. With this fresh set of
volunteers, this did not happen. It is not possible for us to say what exactly was the cause
for the reduction in accuracy. 

\subsubsection{Variability across different phones}
We investigated the variability in accuracy across different testing phones under similar conditions, noting a significant 8\% difference between two Redmi Note 9 Pro Max models of different build years. Additionally, to capture environmental variability, we logged data at the same place but at different times of the day. This resulted in a 9\% difference in accuracy for the same phone and a 13\% difference in accuracy for different phones. Specifically, the Galaxy A53 (2023) achieved the highest average accuracy of 92.3\% in ambient sensing, while the Redmi K20 Pro (2020) exhibited the lowest at 81.8\%. This disparity suggests that older models generally have lower accuracy, primarily due to GPS receiver noise. However, after UKF followed by LDA, the maximum accuracy difference decreased to 3\% for data logged at the same time and 4.2\% for data logged at the same place but at different times. This underscores the effectiveness of our approach in mitigating noisy readings across diverse phone models and handling varying ambient conditions at the same location.
\section{Finding the Indoor Floor Layout}
\label{layout}

Our goal is to create an indoor floor map using semi-processed GPS data. Prior work~\cite{zhou2018graph} has used
WiFi RSSI (received signal strength indicator), user
activities (measured via accelerometers or pressure sensors), and graph optimization
techniques to construct the indoor map. These approaches do not work for us given that the RSSI proves to be
quite a feeble technique in the case of GPS signals. For WiFi the source of the signal is nearby, hence, its
strength variation carries more information. However, for us the source is very far away and the variation observed
is minimal. It is often affected by noise to a much greater degree. This had necessitated the need for UKF and LDA
in the first place. Nevertheless, we use RSS ($\agc$) 
as the starting point of our investigation and then build on it since
is clearly inadequate for our purpose.

\subsection{Study of GPS RSS Patterns}
To study GPS RSSI variations across different areas of a large indoor area, 
we conducted an experiment in a dormitory corridor.
It measured 32.3 m $\times$ 25.6 m (refer to Figure~\ref{fig:floormapandheatmap}(a)). 

\begin{figure}[!htbp]
\vspace{-4ex}
    \hspace*{-1ex}
    \subfloat[]{%
        \includegraphics[scale=0.3, trim=0 0cm 0cm 0cm]{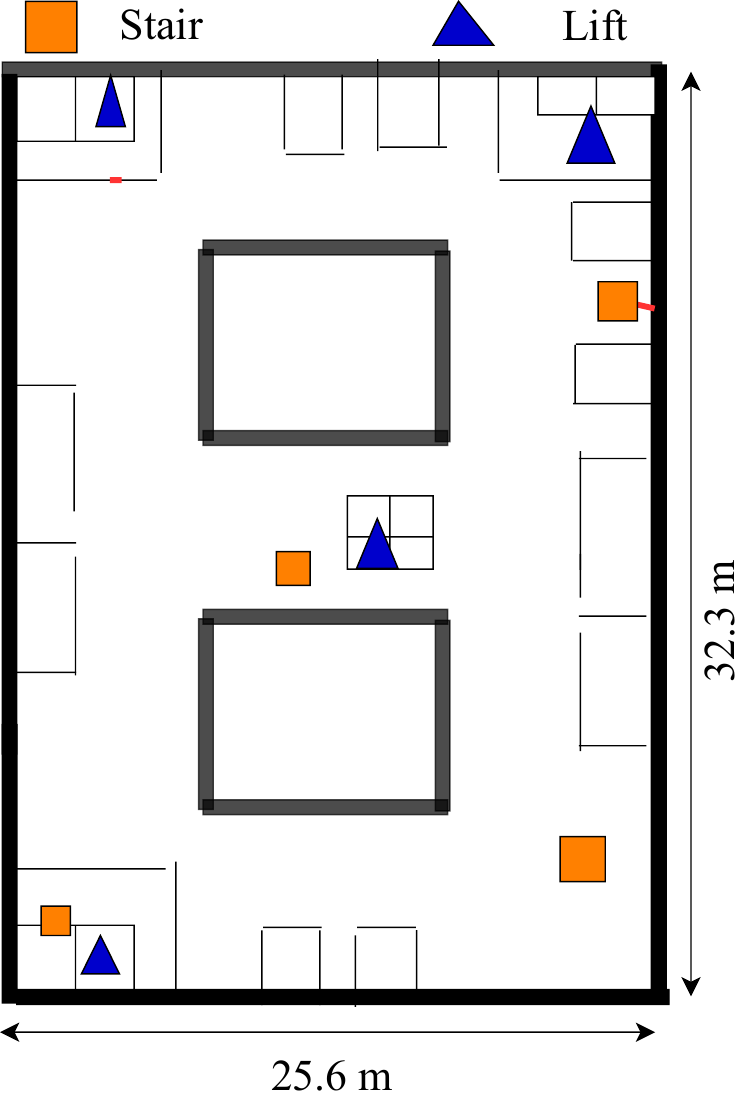}
    }
    \subfloat[]{%
        \hspace{-1ex}
        \raisebox{6.5ex}{
            \includegraphics[scale=0.25, trim=0cm 0cm 0cm 0cm]{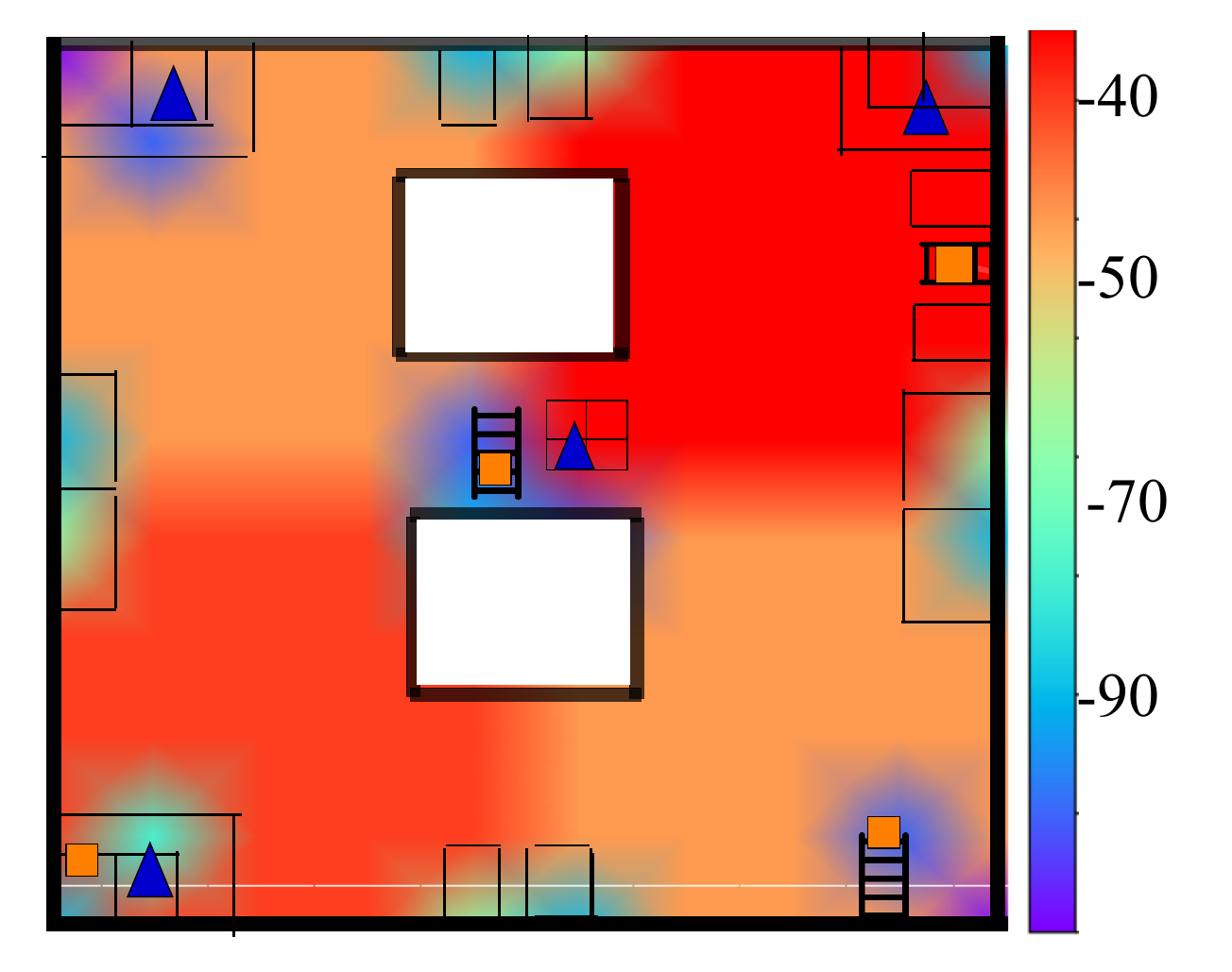}
        }
    }
    \caption{\small {\bf (a)} Floor layout of the dormitory corridor and {\bf (b)} GPS RSS heatmap of the dormitory corridor}
    \label{fig:floormapandheatmap}
\end{figure}

Since there are no GPS-based heatmap simulators available, and civilian GPS operates at a frequency that is very close
to 2.4 GHz WiFi, we utilized NetSpot, a commercial tool designed for WiFi signal strength mapping.
We generated a heatmap depicting the relative GPS RSS levels across the site 
(see Figure~\ref{fig:floormapandheatmap}(b)). We assumed ideal, zero-noise conditions. 
To simulate a GPS setting, we introduced two access points situated outside the site, specifically in the open areas to the northeast and southwest. The differences in the RSS readings indicate their
potential for inferring the layout. We scaled the data to match the real-world GPS signal intensity, which
was measured 
using the RF Explorer Spectrum Analyzer at all points of interest:
near
entrances, lifts, stairs and corners. A correlation of 0.83 was observed, which is quite high. The heatmap clearly identifies the points of interest such as the three lifts at the corners, the rooms and the entrance to the stairs.
But the data still does not tell us what is there at these points of interest. Even with WiFi signals, 
accelerometer and pressure sensor data was required. We are replacing it with semi-processed GPS data and the GPS
location information, which we have not used up till now. 

\subsection{Trajectory Creation Methodology: Use Location Data}

\subsubsection{Activity Landmarking} Our aim is to
identify floor {\em landmarks} such as pathways, stairs, elevators, rooms,
and obstructions based on user activity and GPS-based location information. We will not use data from
any other sensor. For example, users typically walk straight along pathways, turn at
obstructions, and engage in different activities on stairs or in elevators. These activities can be captured using
our human activity recognition (HAR) framework. 
Notably, elevators act as Faraday cages~\cite{Faradaycage2024}, and they block GPS
signals. 
We shall rely on multiple
users to collect this data -- semi-processed GPS parameters, HAR fingerprints and trajectory information.
Then we shall merge these results to {\bf identify the
landmarks}.

\subsubsection{Trajectory Alignment}

We collected the walking trajectories of our volunteers. It is necessary to employ a
trajectory {\em alignment} algorithm (akin to~\cite{zhou2018graph}) because different people will
produce different trajectories. Let us briefly described the algorithm in \cite{zhou2018graph}.
It involves using a transformation matrix, which it creates, for adjusting the curvilinear paths, 
translation and rotation. This matrix is based on common ``activity landmarks'' in the different
trajectories. Trajectory coordinates are classified as either
activity landmark coordinates (ALC) or non-activity landmark coordinates (NALC). 
When merging trajectories, we focus on
common ALCs and calculate their relative coordinates to determine the necessary translations and rotations. This iterative
process ensures a more accurate and consistent overall map by aligning each new path to the existing map.
The advantage of aligning trajectories is that we create a robust skeleton of the internal map. A single observation
is prone to a lot of noise especially when using GPS signals. 

After aligning the trajectories with themselves and with a virtual coordinate system, we apply 
graph optimization techniques (refer to 
~\cite{zhou2018graph}) to further refine the locations of the ALC and NALC trajectory points.
The Levenberg-Marquardt algorithm~\cite{ranganathan2004levenberg} is used for optimization.
The result is a map of the site with the ALC points (specifically) indicated.

\subsection{Evaluation of the Accuracy of Classifying ALC Points}

\subsubsection{Data collection}
10 volunteers participated in the data logging process. The data was collected using the GnssLogger Android
application, the same tool employed for event classification. Each volunteer carried a device with the GnssLogger app
installed and traversed the floor, and sometimes used the services at the landmarks. To ensure robustness and mirror
real-time scenarios, participants were requested to start and end their journeys at random points on the floor.

\subsubsection{Results of the Trajectory Creation Algorithm}

\begin{figure*}[htb]
    \subfloat{
\hspace*{-5ex}
    \includegraphics[scale=0.263, trim ={ 0.5cm 0cm 0cm 0cm}]{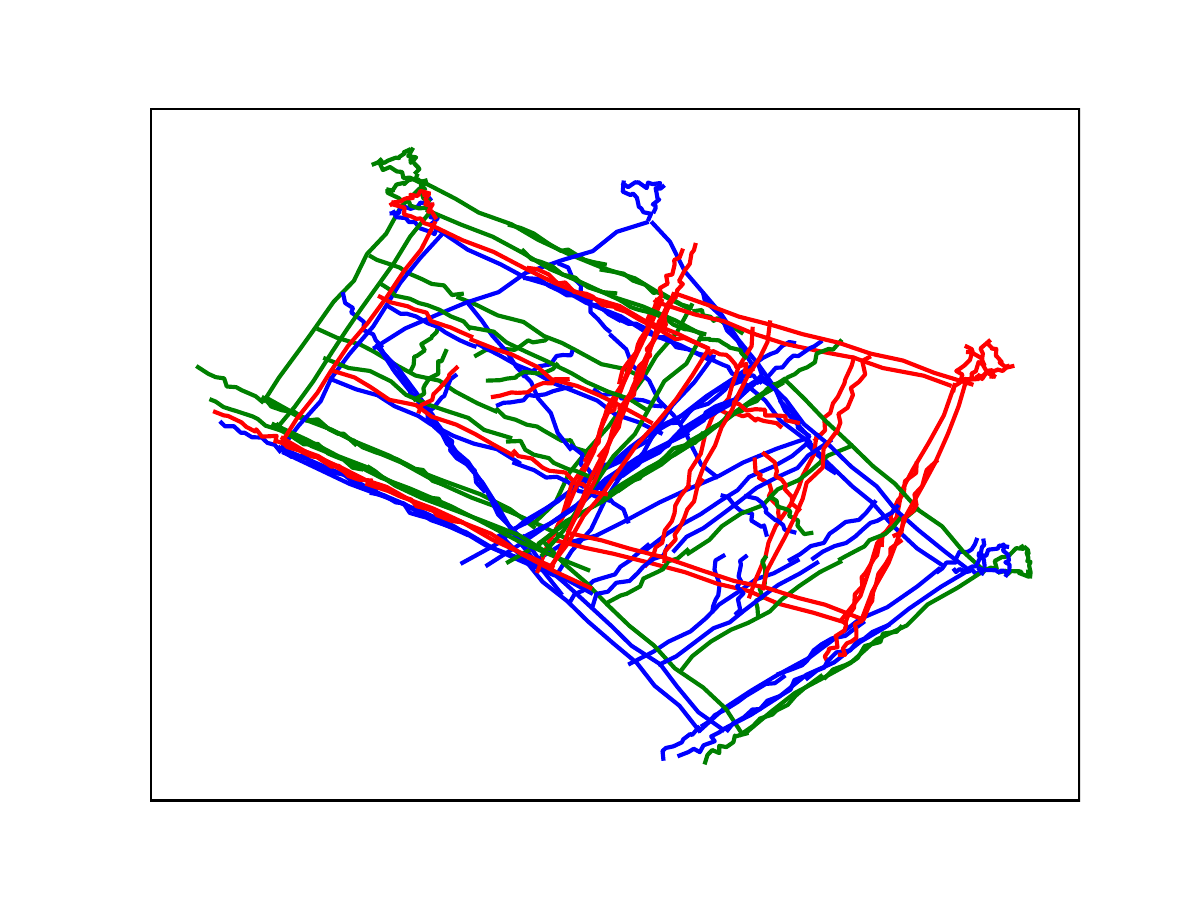}}
    \subfloat{
    
\hspace*{-3ex}
    \includegraphics[scale= 0.32, trim ={ 1cm 0cm 1.5cm 0cm}]{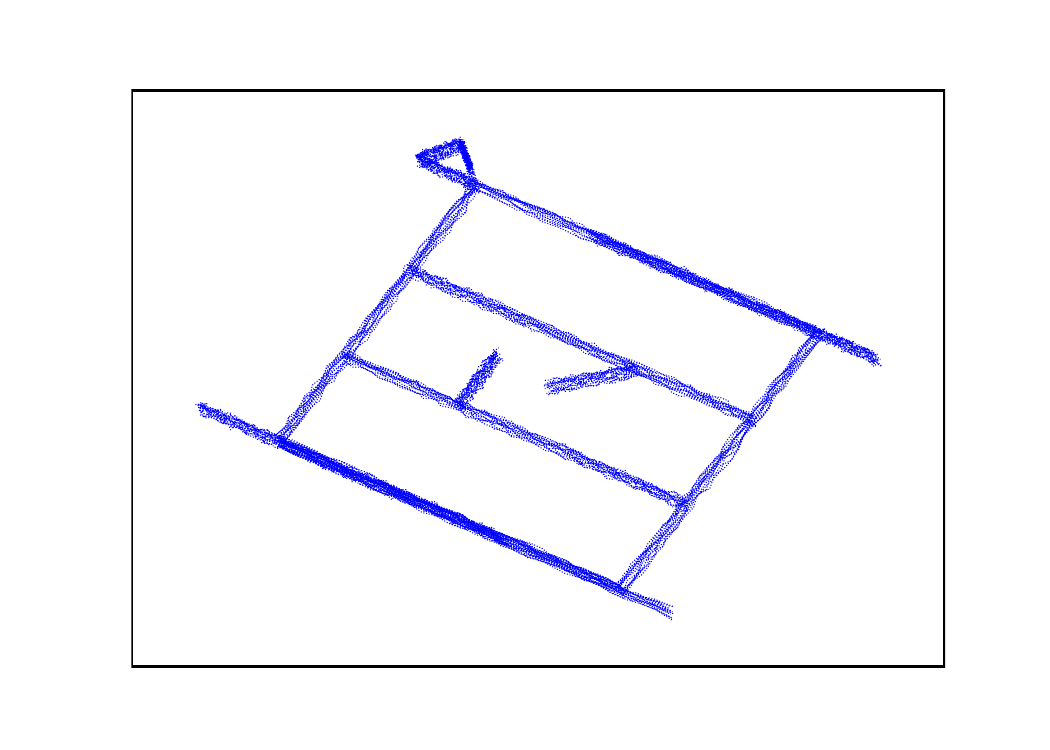}}
        \subfloat{

\hspace*{-2ex}
    \includegraphics[scale= 0.32,trim ={ 0cm 0cm 1.5cm 0cm}]{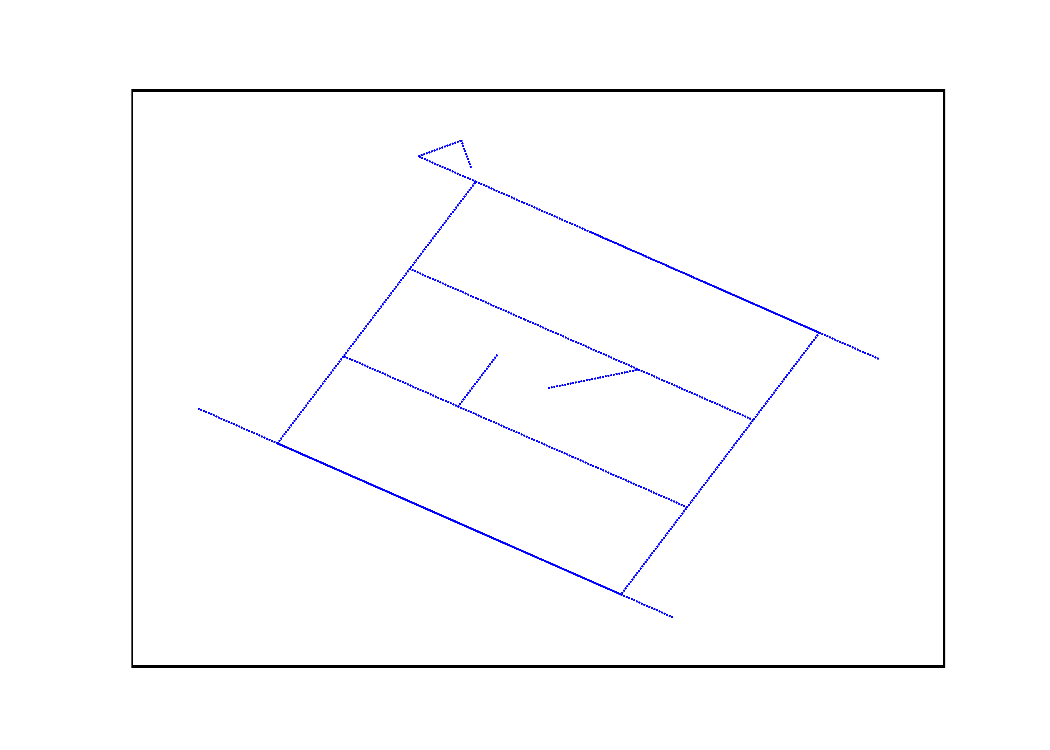}}
        \subfloat{

        \hspace*{-1ex}
        \includegraphics[scale=0.27]{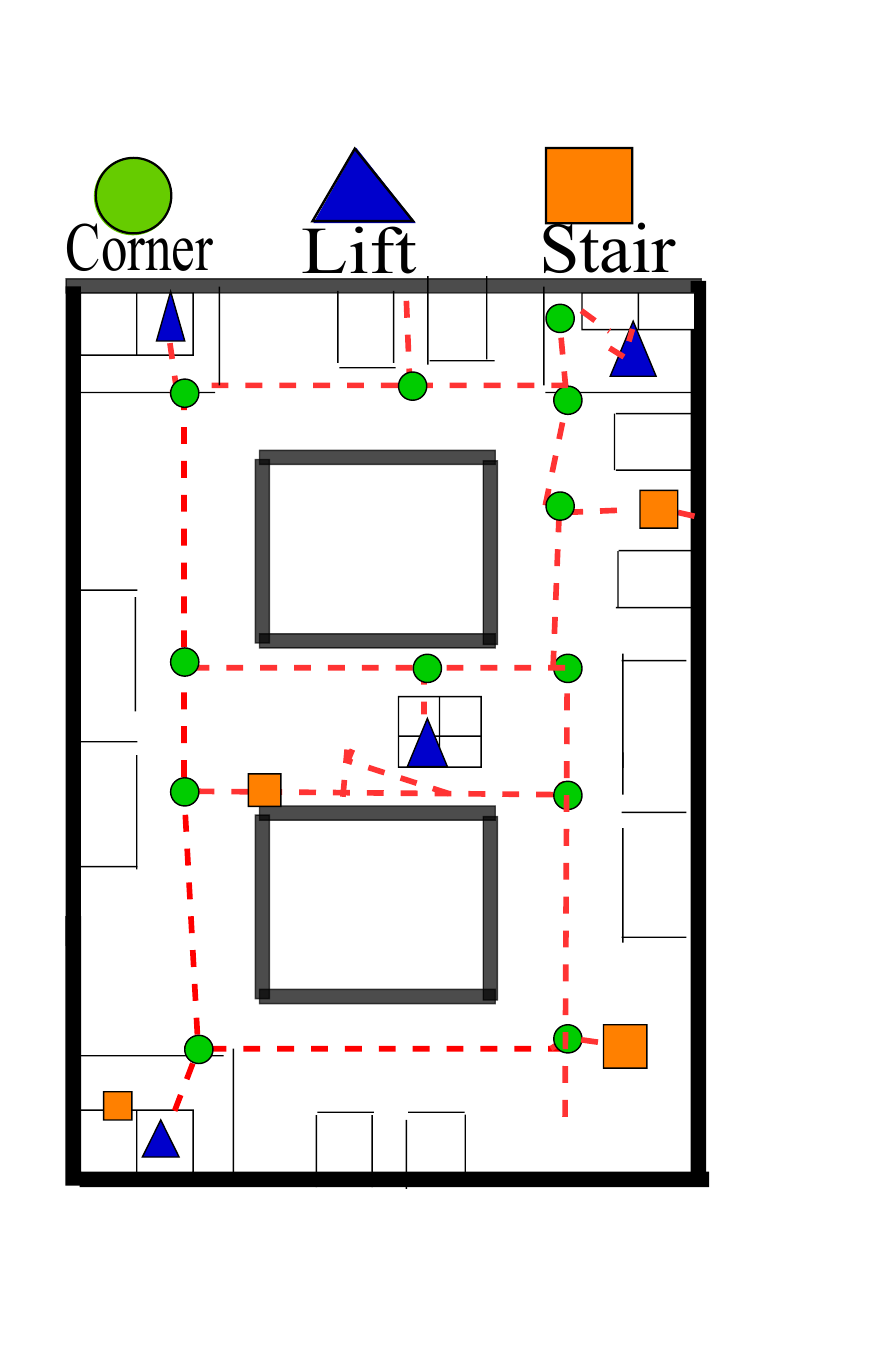}
        }
    \caption{  { Result of the indoor floor mapping experiments: {\bf(a)} Raw trajectories, {\bf (b)} After trajectory alignment}, {\bf (c)} After graph optimization, and {\bf (d)} Final result after aligning with a coordinate system}
    \label{fig:floormapresult}
    \vspace{-2ex}
\end{figure*}

Figure~\ref{fig:floormapresult} 
illustrates the outcome of our mapping process(setting: dorm-room corridor). The activity landmark trajectory data is
shown in Figure \ref{fig:floormapresult}(a). Given the initial random alignment of the trajectories, we realign them
to eliminate inconsistencies as shown in Figure~\ref{fig:floormapresult}(b). The resultant map contains noise,
which we remove using graph optimization
to obtain an optimized floor map with reduced errors (see Figure~\ref{fig:floormapresult}(c)).
The final step involves aligning the generated map with the ALC information. 
The final floor map is shown in
Figure~\ref{fig:floormapresult}(d). Note that we add the real layout in the background only for the purpose
of better visualization. 

\subsubsection{Accuracy Assessment}

\paragraph{\textbf{Layout Shape}} To evaluate the quality of the generated map, we use the following metrics~\cite{zhou2018graph, zhou2015alimc,
shen2013walkie}:

\begin{itemize}
\vspace{-1ex}
    \item  \textbf{Graph Discrepancy Metric (GDM):} This measures the differences between the landmark points of the generated map and the real map (ground truth) using Euclidean distance. A smaller GDM indicates a closer match between the generated map and the real map.
    \item  \textbf{Shape Discrepancy Metric (SDM):} This assesses the differences in the overall shape of the generated
    map as compared to the real map. It involves uniformly sampling points along lines connecting landmarks of both maps
    and measuring the distances between corresponding sampling points. A smaller SDM indicates a greater similarity in
    shape between the two maps.
\end{itemize}

\begin{figure}[!htbp]
\vspace*{-5.5ex}
\hspace*{-1ex}
      \subfloat{%
          \includegraphics[scale=0.185, trim ={ 0 0cm 0cm 0cm}]{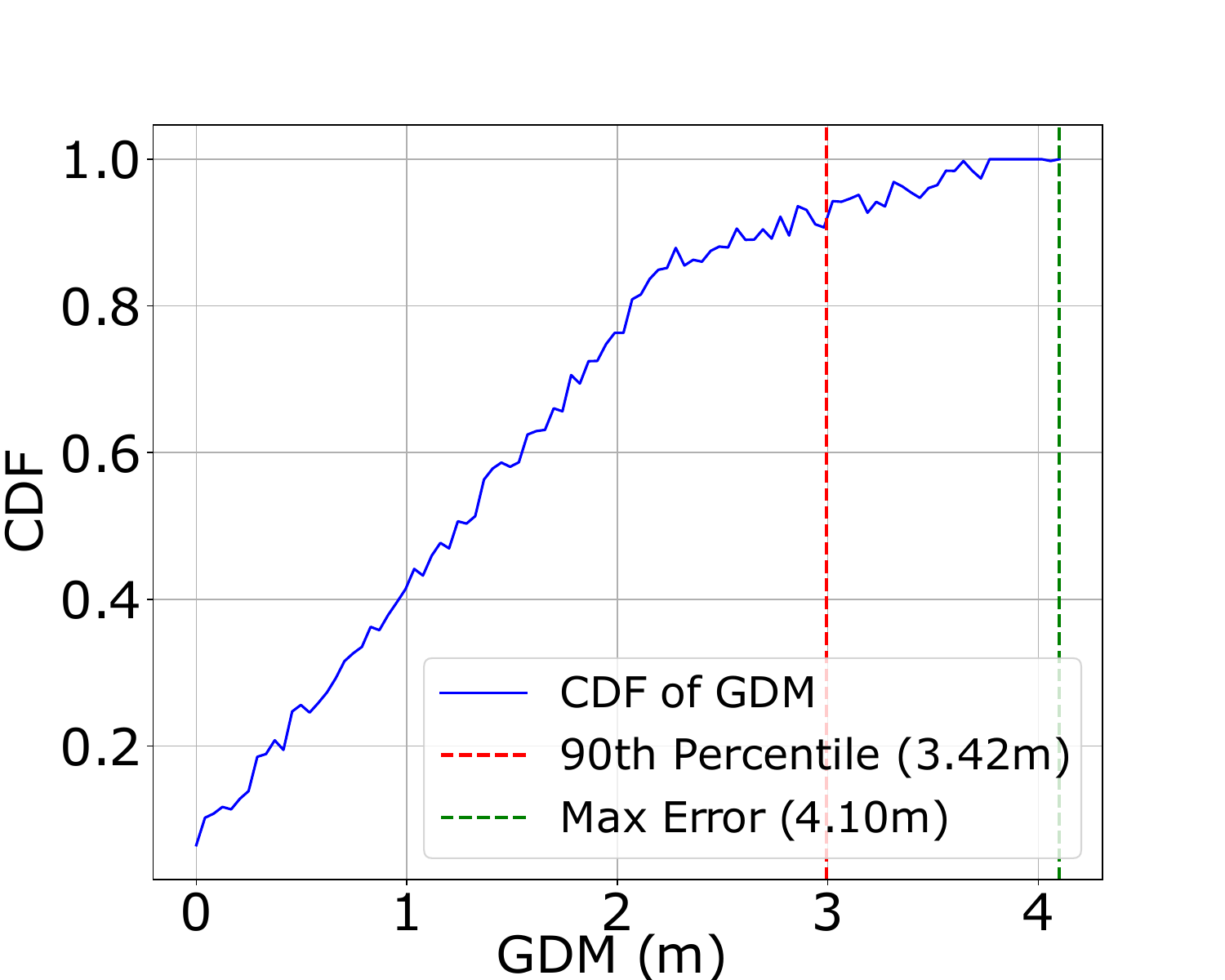}}
      \subfloat{%
\vspace*{-2.5ex}
\hspace*{-3ex}
         \includegraphics[scale=0.185, trim ={ 0cm 0cm 0cm 0cm}]{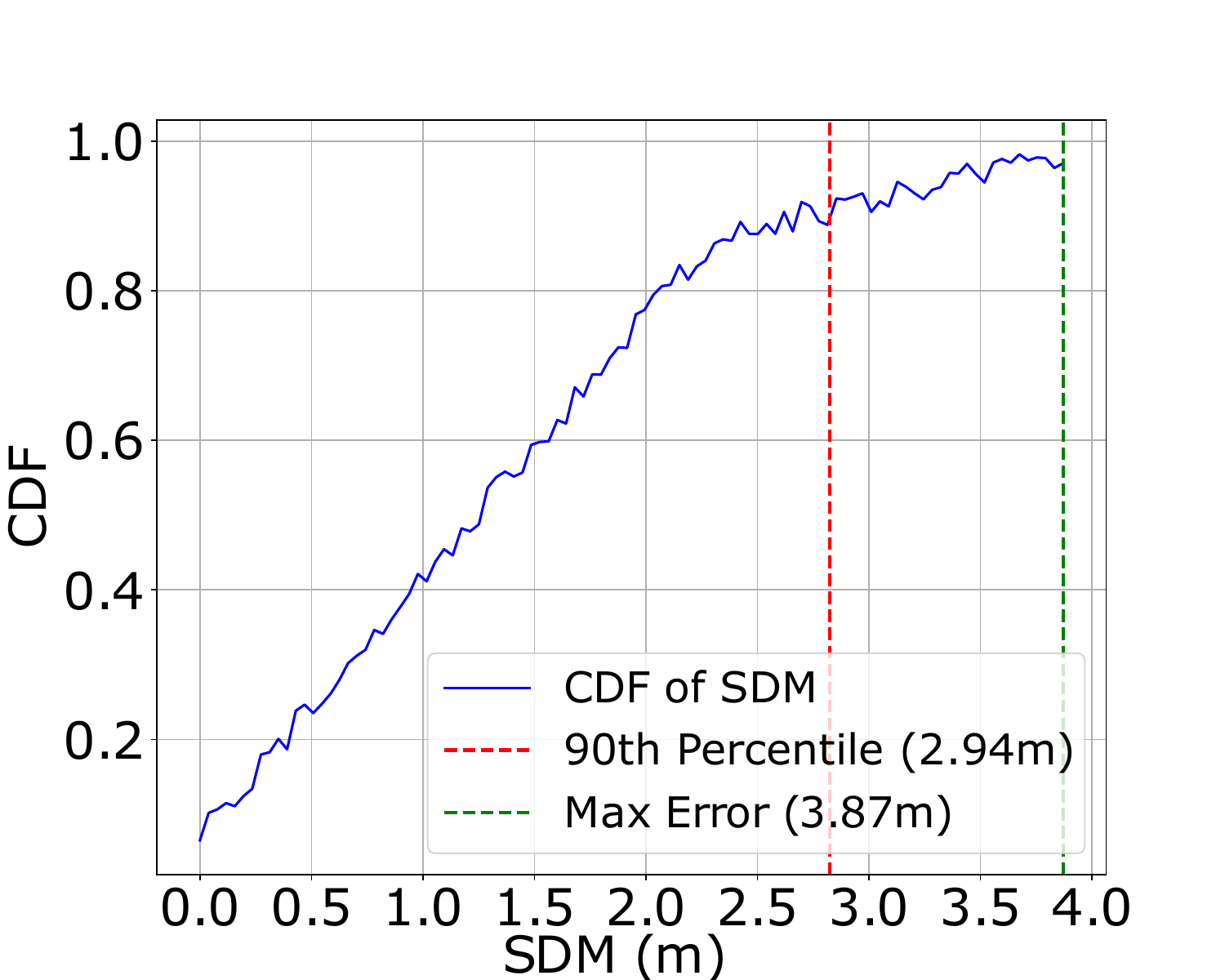}}
\vspace*{-1ex}
  \caption{\small  {CDF of  {\bf(a)} GDM and {\bf (b)} SDM}}
  \label{fig:layoutcompare}
\vspace*{-3ex}
\end{figure}

Figure ~\ref{fig:layoutcompare}(a) shows the cumulative distribution function (CDF) plot for the GDM. The maximum error
observed is 4.10 m, and the $90^{th}$ percentile error is 3.42 m. Similarly, for SDM, the maximum error is 3.41 m, with
the $90^{th}$ percentile error is 2,94 m (refer to Figure~\ref{fig:layoutcompare}(b)). 
Our results align closely with the
related work~\cite{zhou2015alimc, zhou2018graph}, affirming the efficacy of our method in constructing floor maps.

\paragraph{\textbf{Landmark detection accuracy}}
We utilized a GPS-based Human Activity Recognition (HAR) system to accurately detect landmarks categorized as stairs, rooms, lifts, and empty corners within our framework. Table~\ref{tab:landmarkaccuracy} shows the accuracy of the models (RF, DT, GB, KNN, NB, SVM) across four landmarks: Lift, Stairs, Rooms, and Corners.  RF shows strong performance with an average accuracy of 90.15\%. DT and GB also perform well, with average accuracies of 87.5\% and 87.7\% respectively. KNN has a moderate average accuracy of 81\%. NB performs adequately with an average accuracy of 70.8\%, but has a notably low accuracy for Lift (61.4\%). SVM has the lowest average accuracy of 50.2\%, with particularly poor performance for Rooms (54.7\%) and Corners (34.3\%).

\begin{table}[!htb]
\centering
\small{ 
\centering
\caption{Accuracy of activity classifier models}
\label{tab:landmarkaccuracy}
\begin{tabular}{|l|l|l|l|l|}
\hline
\textbf{Classifier} & \textbf{Lift} & \textbf{Stairs} & \textbf{Rooms} & \textbf{Corners} \\ \hline
RF & 98.5 & 96.2 & 86.2 & 79.7 \\ \hline
DT & 96.7 & 93.1 & 77.9 & 82.3 \\ \hline
GB & 96.6 & 93.0 & 78.3 & 82.8 \\ \hline
KNN & 91.8 & 84.4 & 72.9 & 74.9 \\ \hline
NB & 61.4 & 80.7 & 77.8 & 63.2 \\ \hline
SVM & 61.6 & 57.8 & 54.7 & 34.3 \\ \hline
\end{tabular}%
}
\end{table}
\section{Exploitation by Android Application}
\label{sec:invasion}

By exploiting the {\footnotesize \texttt{ACCESS\_PRECISE\_LOCATION}} permission, apps can covertly log semi-processed GPS
data, which as we have seen poses privacy risks. 
While users may agree to precise location access for legitimate services, apps could misuse
data for targeted ads or privacy breaches. Android 12 introduced a privacy option for selecting between fine and
coarse
location permissions. Before this, there was only a single location permission. 
This enhanced the control that users had over
location data access~\cite{finepermission}. 

\noindent{\textbf{\textit{Countermeasures:}}} Android 12 (and beyond) users can limit permissions to {\em coarse location},
thus
restricting app access to the semi-processed GPS data. Android mandates developers to explicitly outline the required
permissions in the app description~\cite{androregulation} file. Android 11 and earlier versions users are often
advised to check
descriptions for fine location requests. They often overlook such details
during installation~\cite{2023policy} risking privacy
breaches. 

\subsection{Impact} Disabling fine location permissions isn't practical, as developers may justify its need for
location-based services.  Even vigilant users regardless of
whether they check app descriptions for requested permissions or choose
between permissions are still vulnerable to privacy breaches. This vulnerability persists as users are unaware of the
privacy risks associated with logging semi-processed GPS data, owing to the \textbf{\textit{absence}} of 
relevant disclosures in the
official Android documentation~\cite{finepermission}.

Additionally, devices running Android 10 or later must support semi-processed GPS measurements, while for Android 9 and
earlier, this is mandatory for devices manufactured in 2016 or later~\cite{rawgnssandoid}. Currently,
\textbf{\textit{over 90\%}} of Android phones support semi-processed measurements~\cite{rawgnssandoid}, exposing a
substantial portion of the user base to this kind of a covert attack.
\section{Related Work}
\label{related}

Given that there is an overlap in the prior work on activity and ambience recognition,
we combine them and discuss the combined related work in Section~\ref{sec:ambharrel}.
Subsequently, we shall
discuss the related work on layout detection in Section~\ref{sec:locrel}. 

\subsection{Ambient Sensing and HAR} 
\label{sec:ambharrel}
\begin{table*}[h!]
\centering
\begin{tabular}{|l|l|l|c|c|c|c|c|c|}
\hline
\rowcolor{gray!30} \textbf{Year} & \textbf{Work} & \textbf{Approach} & \textbf{WiFi} & \textbf{IMU} & \textbf{Cell Tower} & \textbf{GPS} & \textbf{RFID} & \textbf{Camera} \\ \hline \hline
2021 & Ramirez et al.~\cite{ramirez2021fall} & Video-based trajectory & \xmark & \xmark & \xmark & \xmark & \xmark & \cmark \\ \hline
2016 & Wang et al.~\cite{wang2016device} & Wireless signal CSI patterns & \cmark & \xmark & \xmark & \xmark & \xmark &  \xmark\\ \hline
2017 & Gao et al.~\cite{gao2017csi} & Wireless signal CSI patterns & \cmark & \xmark & \xmark & \xmark & \xmark & \xmark \\ \hline
2020 & Muaaz et al.~\cite{muaaz2020wiwehar} & WiFi CSI + IMU sensors & \cmark & \cmark & \xmark & \xmark & \xmark & \xmark \\ \hline
2015 & PAWS~\cite{gu2015paws} & RSS-Based & \cmark & \xmark & \xmark & \xmark & \xmark & \xmark \\ \hline
2020 & Bhat et al.~\cite{bhat2020human} & WiFi and RSS-based & \cmark & \xmark & \xmark & \xmark & \xmark & \xmark \\ \hline
2020 & Shuaeib et al.~\cite{shuaieb2020rfid} & RFID-Based & \xmark & \xmark & \xmark & \xmark & \cmark & \xmark \\ \hline
2021 & Sekiguchi et al.~\cite{sekiguchi2021phased} & GPS coordinates + Other wireless signals & \cmark & \xmark & \cmark & \cmark & \xmark & \xmark \\ \hline
2020 & Bui et al.~\cite{bui2020gps} & processed GPS signals  & \xmark & \xmark & \xmark & \cmark & \xmark & \xmark \\ \hline
2024 & Zhu et al.~\cite{zhu2024deep} & processed GPS signals & \xmark & \xmark & \xmark & \cmark & \xmark & \xmark \\ \hline
2024 & \textbf{\emph{AndroCon}} & Semi-processed GPS signal parameters & \xmark & \xmark & \xmark & \cmark & \xmark & \xmark \\ \hline
\end{tabular}
\caption{Comparison of Event Classification Approaches}
\label{table:eventclassification}
\end{table*}

We present a brief comparison of the related work in Table~\ref{table:eventclassification}. Image-based solutions are prominent due to their
high accuracy~\cite{ramirez2021fall}. However, their widespread adoption is
limited by the high cost of processing image data and security concerns. 

In response, wireless signal patterns:  RSS and channel state information (CSI) is used for ambient and activity recognition. Wang et al.~\cite{wang2016device} introduced a deep learning
technique, a sparse autoencoder (SAE), which 
recognizes events using WiFi CSI signals. Gao et
al.~\cite{gao2017csi} developed another CSI-based system for event
classification using deep learning, where they converted
CSI measurements into radio images, extracted features, and then applied an SAE network for better accuracy. Muaaz et
al.~\cite{muaaz2020wiwehar} proposed WiWeHAR, a multimodal HAR system that combines WiFi 
CSI data and wearable inertial
measurement unit (IMU) sensor data. These systems demand
precise feature engineering and efficient signal processing and noise removal. 

RSS-based solutions are comparatively more noise-tolerant.
One of the earliest works is PAWS~\cite{gu2015paws}, which
utilizes ambient WiFi signals to create RSSI fingerprints.
Bhat et al.~\cite{bhat2020human} refined existing techniques to develop
a recognition system using the
RSS of a single communication channel. 
Shuaeib et al.~\cite{shuaieb2020rfid} introduced an RFID-based indoor HAR system
that uses RSS from passive RFID tags to track activity in real-time 
by mapping the analyzed data to reference datasets.
Such solutions rely heavily on the availability of good-quality WiFi signals,
which may not always be feasible especially in outdoor settings. GPS, on the other
hand, is more ubiquitously available. 

Sekiguchi et al.~\cite{sekiguchi2021phased} 
categorize events based on GPS coordinates along with cell tower and WiFi
signal data. Our approach does not use GPS coordinates or any other kind of location information.
We solely focus on semi-processed GPS signal
parameters, which are almost always available. We do not use any WiFi or assisted-GPS
signals (signals from cellphone towers). 

Bui et al.~\cite{bui2020gps} utiized the magnitude of the received GPS signal for distinguishing between indoor and outdoor environments. Zhu et al.~\cite{zhu2024deep} classify ambient conditions based on visible satellites, GNSS distribution, C/N0, and multipath effects. However, both studies utilize processed signal data, which yields less accurate results compared to our research. Our work also encompasses diverse settings that were not accounted for in either study.

\subsection{Indoor Layout Mapping}
\label{sec:locrel}

\begin{table*}[!htb]
\centering
\begin{tabular}{|c|l|p{5cm}|c|c|p{2.4cm}|c|c|}
\hline
\rowcolor{gray!30} \textbf{Year} & \textbf{Work} & \textbf{Approach} & \textbf{SLAM} & \textbf{WiFi} & \textbf{Smartphone sensors} & \textbf{IMU} & \textbf{GPS} \\ \hline \hline
2020 & Karam et al.~\cite{karam2020strategies} & IMU + LiDAR & \cmark & \xmark & \xmark & \cmark & \xmark \\ \hline
2012 & Alzantot et al.~\cite{alzantot2012crowdinside} & User trajectories & \xmark & \xmark & \cmark & \xmark & \xmark\\ \hline
2013 & Shen et al.~\cite{shen2013walkie} & WiFi RSS alignment & \xmark & \cmark & \xmark & \xmark & \xmark \\ \hline
2016 & Gu et al.~\cite{gu2016using} & User trajectory +WiFi+Bluetooth & \xmark & \cmark & \cmark & \xmark & \xmark \\ \hline
2014 & Philipp et al.~\cite{philipp2014mapgenie} & Pedestrian movement+crowd-sourced structural (external) information of buildings & \xmark & \xmark & \xmark & \cmark & \xmark \\ \hline
2015 & Zhou et al.~\cite{zhou2015alimc} & Link-node approach & \xmark & \cmark & \cmark  & \xmark & \xmark \\ \hline
2018 & Zhou et al.~\cite{zhou2018graph} & Link-node approach+GO & \xmark & \cmark & \cmark  & \xmark & \xmark \\ \hline
2024 & \textbf{AndroCon} & GPS Semi-processed data + user trajectory & \xmark & \xmark & 
 \xmark & \xmark & \cmark \\ \hline
\end{tabular}
\caption{Comparison of Indoor Floor Mapping Approaches}
\label{table:indoormapping}
\end{table*}

We show a comparison of related studies in Table~\ref{table:indoormapping}.
Most approaches utilize
Simultaneous Localization and Mapping (SLAM), a computer vision
technique designed to localize a robot in an unknown environment
while concurrently generating a map. 
Conventional SLAM techniques typically depend on visual cues such as landmarks,
camera-detected obstacles, sonar data 
or laser-range sensors (Kamar et al.~\cite{karam2020strategies})~\cite{lu2020see, plikynas2020indoor, kang2020review}. However, these approaches may incur high costs or maybe
quite intrusive in terms of privacy.

Alzantot et al.~\cite{alzantot2012crowdinside} propose to 
use crowd-sourced user trajectories to construct indoor
maps. However, this method was susceptible to errors due to the non-alignment of trajectories. Adressing the issue, Shen et al.~\cite{shen2013walkie} present an indoor pathway mapping system using WiFi RSS fingerprints to align trajectories. Gu et al.~\cite{gu2016using, gu2017wifi} further improved the performance by employing Bluetooth and WiFi data for trajectory alignment, but sparse WiFi access point
deployment poses practical challenges. Philipp et al.~\cite{philipp2014mapgenie} propose Mapgenie -- generating maps from
pedestrian movement traces and utilizing external building 
data and applies grammatical rules to model and interpret pedestrian movement patterns; however, it
requires a foot-mounted IMU that is typically not found on smartphones. Zhou et al.~\cite{zhou2015alimc}  use a link-node approach (landmarks as node and pathways and links). This method utilizes smartphone sensors and WiFi RSS fingerprinting, complemented by geometric scaling methods. However, challenges arise in accurately representing curved features within the maps. Zhou et al. \cite{zhou2018graph} address these challenges through graph optimization methods.

Our approach draws inspiration from Zhou et al.~\cite{zhou2018graph} that
leverages WiFi RSS and MAC address fingerprinting
for trajectory alignment and landmark identification. 
Sadly, Zhou et al.'s method relies heavily on mobile sensors
such as gyroscopes and compass readings, 
which may introduce errors and reduce reliability. Moreover, it requires
extensive user permissions and depends on the availability of WiFi access points.
In contrast, our proposed method
utilizes semi-processed GPS data and does not need any other information.

For both event recognition and indoor floor mapping,
image-based solutions\cite{ramirez2021fall, kang2020review} captured using cameras are highly effective; however,
they raise significant
privacy concerns. This has limited their widespread adoption. 

In contrast, methods utilizing EM waves such as WiFi, Bluetooth or GPS
are preferred due to their
covert nature. Many of these methods have matured over time and they provide solutions
with a comparable quality as camera-based designs do. Our solution, AndroCon, also
provides similar accuracies (if not better).
\section{Conclusion}
\label{sec:conclusion}

The fact that semi-processed GPS data can be used to sense the ambient, recognize human activity
and figure out floor layouts was hitherto unknown. We successfully showed that
all of the above can be achieved, and that too with a high accuracy that ranges from roughly
99.5\% in controlled conditions to 
87\% in absolutely uncontrolled conditions. We conducted an extensive set of experiments with
tens of volunteers, many phones of different makes and brands, diverse set of 
scenarios and thousands
of sample points. We collected data for a year across a large geographical area -- some
of the collection points were 1000 kms away from the place where this research was carried out.
We also collected data on flights, cruise ships and high-altitude locations. Our results are
thus robust.

Furthermore, our approach can construct floor maps with a
maximum error margin of 4.1m as compared to the ground truth.
We can classify points of interest within an indoor layout such as elevators, stairs,
corridors, empty corners and rooms with a roughly 90.15\% accuracy, while just relying
on GPS data. 
Currently, Android does not address this vulnerability, which leaves 
approximately 90\% of users exposed.
\section{Responsible Disclosure}

We reported the identified vulnerability and use-cases,
excluding the floor mapping use-case to the Android security team. They were able to reproduce
some of the results and they
acknowledged our concern.  

\bibliographystyle{IEEEtran}
\bibliography{references}

\end{document}